\documentclass{article}

\usepackage{subcaption}
\usepackage{etoolbox}
\usepackage{lineno}
\usepackage{enumitem}

\usepackage{amssymb}

\usepackage[toc,page]{appendix}

\usepackage{floatrow}
\usepackage{dsfont}\usepackage[margin=1in]{geometry}

\usepackage{colortbl}

\usepackage{amsmath, amsthm, amsfonts}
\usepackage[mathscr]{eucal}
\usepackage{bbm}
 \usepackage{setspace}

\usepackage{xcolor}

\usepackage{graphicx} 
\graphicspath{{./Figures/}}

\usepackage{lscape}
\usepackage{url}

\usepackage{rotating}

\usepackage[numbers,sort&compress]{natbib}
\usepackage{filecontents}

\providecommand{\keywords}[1]
{
  \small	
  \textbf{Keywords:} #1
}

\newlist{outerenum}{enumerate}{1}
\setlist[outerenum]{label={\arabic*.}}

\newcommand\notype[1]{\unskip}

\title{Behavior Associations in Lone Actor Terrorists~\thanks{This material is based upon work supported by the National Science Foundation (Grant  No.1901721)}}
\author{Ayca Altay$^{a,*}$, Melike Baykal-G\"ursoy$^a$, Pernille Hemmer$^b$\\
  \small  $^a$Industrial \& Systems Engineering, Rutgers University\\
  \small $^b$Department of Psychology, Rutgers University\\
  \small $^*$ayca.altay@rutgers.edu
}

\begin{document}
\maketitle

\abstract{Terrorist attacks carried out by individuals or single cells have significantly accelerated over the last 20 years. This type of terrorism, defined as lone-actor (LA) terrorism, stands as one of the greatest security threats of our time. Research on LA behavior and characteristics has emerged and accelerated over the last decade.  While these studies have produced valuable information on demographics, behavior, classifications, and warning signs, the relationship among these characters are yet to be addressed. Moreover, the means of radicalization and attacking have changed over decades. This study first identifies 25 binary behavioral characteristics of LAs and analyzes 192 LAs recorded on three different databases. Next, the classification is carried out according to first ideology, then to incident scene behavior via a virtual attacker-defender game, and, finally, according to the clusters obtained from the data. In addition, within each class, statistically significant associations and temporal relations are extracted using the \textit{A-priori} algorithm. These associations would be instrumental in identifying the attacker type and intervene at the right time. The results indicate that while pre-9/11 LAs  were mostly radicalized by the people in their environment, post-9/11 LAs are more diverse. Furthermore, the association chains for different LA types present unique characteristic pathways to violence and after-attack behavior.


\section{Introduction}

Terrorism is defined as an act of aggression or violence against noncombatants with the objective of affecting policy makers indirectly by intimidating the target audience~\cite{phillips2011lone}. Terrorist threat does not arise from a desire for financial gain or mere personal vengeance; there is a higher political, ideological or religious motivation~\cite{spaaij2010enigma}.  Partly due to the success in counterterrorism efforts, the face of terrorism has changed dramatically in recent years. Attacks by groups with defined chains of command have declined, as the prevalence of autonomous cells and individuals has grown. The core focus of this paper, lone actor (LA) terrorism has increased by 134\% over the last 20 years~\cite{phillips2011lone}. Recent examples include the 2017 Las Vegas mass shooting of concertgoers~\cite{hinksonj} and the 2019 El Paso shooting in a supermarket~\cite{quek2019paso}. 

Even though the definition of ``LA terrorism'' is open to discussion, particularly regarding whether dyads or triads, and members of extremist/terrorist organizations should be included~\cite{spaaij2010enigma, gill2014bombing}, existing literature largely agrees on several characteristics. Spaaij~\cite{spaaij2010enigma} argues that LA attacks are consequences of a personal grudge channeled into a higher cause. As a result, personal and ideological motivations may not be entirely distinguishable. Moreover, LA attacks are rarely impulsive or sudden~\cite{gill2014bombing} and LAs are not as well-organized as terrorist groups, hence, an intended attack preparation usually takes longer than an attack by terrorist groups~\cite{brynielsson2013harvesting}. Finally, LAs may align themselves with extreme movements~\cite{pantucci2011topology}; in fact, they mostly appear to be too extremist even for terrorist organizations~\cite{bakker2011preventing}. The literature also agrees that LA attacks differ from common crime or other assassinations by involving an ulterior political, religious or other ideological component~\cite{brynielsson2013harvesting}; they often come up with their own ideologies that is a mixture of personal vendettas and ideological grievances. In the following subsections, we preview the common characteristics, and` classification structures of LAs and review the prior work on quantitative analyses in this field.

\subsection{Characteristics and Detection Challenges of LAs}

When investigating the characteristics of LAs, the research focuses on demographic, economic, and psychographic characteristics. These characteristics are composed of $(i)$ certain events or incidents that might cause radicalization, $(ii)$ observable actions an LA might commit, and $(iii)$ the environmental response to these observable actions. Some of these characteristics involve events prior to an attack intent, such as having relations with extremist people or mental illness leading to radicalization~\cite{meloy2011role}. Some events or incidents may provoke the idea of an attack and act as a triggering mechanism that can range from a personal hardship such as being fired from a job to a mass incident such as an international conflict between governments~\cite{bakker2011preventing,gartenstein2015radicalization,spaaij2011understanding}. Other behavioral characteristics involve details in attack planning such as the target or weapon selection. 
 
 Distinguishing LAs from people with extremist ideological views constitutes a challenge as the majority of people with extremist ideological views do not pose a security threat~\cite{bakker2011preventing}. In that context, McCauley and Moskalenko~\cite{maccauley2008mechanisms} proposed two radicalization pyramids, one of opinions and one of actions. The ``opinion radicalization pyramid'' has a range from ``neutral'' to ``ideology as a moral obligation'' and the ``action radicalization pyramid'' has a range from ``inert'' to ``terrorist''. According to this study, there is a jump even between the pinnacle of the opinion pyramid (defending ideology as a moral obligation) and the radical action, and this transition is not inevitable. However, the authors also conclude that not all LAs need to climb the opinion pyramid before committing an attack; it is a matter of opportunity rather than an obligation~\cite{maccauley2008mechanisms}. A more extensive behavioral study conducted by Meloy et al.~\cite{meloy2011role} defined eight ``proximal warning behaviors'' for an LA attack. It is essential to have a ``pathway behavior'' from grievance to attack. An LA fixates on a target or an ideology more intensely over time and identifies himself with the cause or a role model. LAs may leak their intent directly or indirectly throughout their pathway. An attack requires familiarity with weapons; hence, they conduct dry runs to test their abilities. Before an attack, there is an ``energy burst'' indicating that the settlements and preparations for the attack have started. Occasionally, LAs directly threaten the target and they may also hint that the attack is their only option or ``last resort''. These ``proximal warning behaviors'' can, then, be used in the Terrorist Radicalization Assessment Protocol (TRAP-18), which is a professional LA risk-assessment framework~\cite{meloy2017trap}. Beside these ``proximal warning behaviors'', TRAP-18 involves 10 ``distal characteristics''. These distal characteristics are $a)$ having a personal grievance and moral outage; $b)$ having been framed by an ideology; $c)$ failing to affiliate with an extremist group; $d)$ depending on virtual communities; $e)$ thwarting of occupational goals; $f)$ changes in thinking and emotion; $g)$ failing of sexual-intimate pairing; $h)$ having a mental disorder; $i)$ being creative and innovative; $j)$ having a criminal violence history~\cite{meloy2017trap}.

Some of these elements are difficult to trace in real life; most physical moves of an LA are hardly traceable. Moreover, conventional attack prevention techniques, such as infiltration or wiretapping, are not effective for LA attacks due to the absence of a group~\cite{brynielsson2013harvesting}. The silver lining in LA detection is that they are commonly radicalized by Internet exchanges; and, therefore leave their ``writeprints''~\cite{meloy2011role,brynielsson2013harvesting}. They spread their views and opinions before committing an actual attack~\cite{brynielsson2013harvesting}. The biggest challenge of detecting an LA online is that search engines cannot access to the ``deep web'' in which such exchanges often take place. 

\subsection{Classification of LAs}

Although, LAs have some commonalities such as acting alone or having fixated on an ideology, their behaviors along the pathway to violence may vary among different attacker types. Consequently, the existing literature proposes different classification domains for LAs. Some examples of these domains are location, purpose, type of target, goals, and ``role in protection''~\cite{bates2012dancing}.

 Pantucci~\cite{pantucci2011topology} categorized LAs as the loner, the lone wolf, the lone wolf pack, and the lone attacker depending on the number of people involved and the existence of a chain of command. Bates et al.'s classification used four dimensions to group LAs: degree of self-radicalization, risk-awareness level of the LA, altruistic motivation, and the number of attacks intended~\cite{bates2012dancing}. Gill et al.~\cite{gill2014bombing} classified LAs in terms of their ideologies: right-wing, islamists, and single issue. They argued that even though LAs as a whole do not have any distinguishing characteristics except mostly being males, subgroups have very noticeable characteristics in terms of demographics, network connectivity and operational success. This ideology-based classification was commonly applied in literature. Meloy and Gill~\cite{meloy2016lone} applied statistical analyses that compared three ideological groups in terms of their distal characteristics and warning behaviors. They found out that right-wing LAs are less fixated on their ideologies, single-issue LAs are less dependent on virtual communities. Sawyer and Hienz~\cite{sawyer2017what} included far-left and hybrid (combination of different ideologies) to their classification, albeit they were less in number compared to the three main ideological groups.

\subsection {Quantitative Approaches for LA Terrorism}

Meaningful statistical analyses over LAs are difficult to conduct since LA attacks are black-swan events that are rare even in extremist ideologies, which makes them impossible to label, let alone forecasted~\cite{bakker2011preventing}. Similarly, commonly applied terrorist social network analysis methods fail due to the absence of a group network~\cite{perliger2011social,li2018terrorist}. However, with the help of social media and online forums, identification, leakage and fixation are traceable online. Brynielsson et al.~\cite{brynielsson2013harvesting} extracted the attack intent of an LA through text mining. Intent appears as an important indicator of an attack in the economic analysis of an LA attack~\cite{spaaij2010enigma,phillips2011lone}. Phillips~\cite{phillips2011lone} implemented a quantitative model that shows a decrease in the probability of an attack in case of higher-level of security (the deterrence effect) but an increase in the probability of an improvised attack plan by momentarily switching the target (the substitute effect). In addition, he modeled the increase in the probability of an attack with the increased financial resources of the attacker (the endowment effect) and embedded the risk-averse/risk-seeking behavior of the attacker in the model (the preference effect) in a game-theoretical scheme. Gordon et al.~\cite{gordon2017potential}  worked on weights of different criteria in the intent of an attacker. They also implemented a Delphi Method, in which a facilitator gathers expert opinions anonymously with the sole purpose of arriving to a consensus by supplying recursive feedback.

Most quantitative studies on LA behavior are composed of summary statistics and hypothesis tests over LA demographics and behavior. A recent study by Philips (2017) compares the USA and other 15 developed countries in terms of casualties and demonstrates that LA attacks are deadly threats especially in the United States. Schuurman et al.~\cite{schuurman2018lone} analyzed the LA cases in North America and Europe between 1985-2015 and provided striking results. They found that over 80\% of LA attacks on American targets between 1999 and 2009 were prevented by the attention of law enforcement or the general public such as neighborhood watches. According to their data, 46\% of LAs had shown violent behavior, 62\% of them had contacts with radical, extremists people or terrorists and 31\% were well-known members of these groups. 86\% of LAs communicated their attack-related ideas via oral communication or social media leakage. Ellis et al.~\cite{ellis2017lone} found that weapons training increases the number of casualties per attack from an average of 1.47 to 2.29. A study by Gill et al.~\cite{gill2014bombing} used hypothesis testing and showed that LAs do not have a defining characteristic as a whole but that the subgroups of right-wing, single issue and jihadist terrorists have their own demographics and network structures. Even within the same subgoup the weapon choice, and the attack preparation of an LA may present different casualty  outcomes. Thus it is clear that although ideology-based classification leads up to common demographic characteristics~\cite{gill2014bombing}, it may not necessarily provide commonalities in terms of behavioral characteristics or responses. 

Another challenge for behavioral analyses is that the term ``behavior'' have different scopes in literature. Besides, the terms ``terrorist behavior'' or ``attacker behavior'' can refer to different concepts. In some studies, the term involves attack-related behavior such as identification or fixation~\cite{brynielsson2013harvesting}, while in some others, ``behavior'' refers to the response, mood or observable actions of a terrorist~\cite{post1985individual}. 

These studies indicate that while knowledge on LA definition, typology, and demographics is available to a certain extent, relationships among these characteristics are yet to be discovered. This paper intends to address this problem and makes the following contributions to the literature:

\begin{itemize}
    \item Provides clarification on the ``behavior'' term with a temporal perspective. 
    \item Identifies 25 binary attributes for LA characteristics based on the data prepared by Hamm and Spaaij~\cite{hammandspaaij}. 
    \item Defines behavior-based attacker types using these binary attributes.
    \item Compares pre-9/11 LAs to post-9/11 LAs to reflect the temporal changes in LA terrorism.
    \item Analyzes the implications among LA behaviors for each type.
    \item Forms chain rules that sum up the evolution process of each attacker type. 
    \item Extracts temporal relations between LA behavior milestones for each type. 
\end{itemize}

The structure of this paper is as follows. Section \ref{third} provides detailed information on the means of LA data gathering and processing, the identification of 25 attributes for LA characteristics, and various LA classification schemes. Section \ref{methodology} introduces the \textit{A-priori}  algorithm, describes its statistical properties and the rule-chain-formation procedure.  Section \ref{overall} provides associations for the pooled database, and Section \ref{temporals} compares pre- and post-9/11 era to demonstrate how LA terrorism has changed through decades. Section \ref{motivate} introduces LA types based on three classification schemes: $i)$ ideology-based, $ii)$ incident-scene-based, and $iii)$ behavioral-based. Section \ref{chains} forms behavioral and chronological association chains of LAs for each type. Section \ref{tempAs} demonstrates the temporal relationship among observable landmarks on the attack pathway. The concluding remarks follow in the last section.

\section{LA Characteristics, Behavior, Data, and Classifications}\label{third}

\subsection{Data for LA Behavior and Characteristics}\label{dataset}

The National Criminal Justice Reference System database prepared by Hamm and Spaaij~\cite{hammandspaaij} involves  98 LA attacks in the USA between 1940-2013. This dataset can be used to obtain data on LA behavior and attack characteristics. While providing extremely valuable data, the database lacks the recent data on LA attacks, given the fact that LA terrorism has been on the rise especially over the last decade. However, the Global Terrorism Database (GTD)~\cite{gtd} holds the records of all terrorist events worldwide and the Mother Jones database~\cite{motherjones} holds the records of all mass shootings in the USA. The GTD records the data of attacks until the end of 2017, and it identifies 192 incidents in the USA between 2014-2017. However, these incidents do not only involve LA attacks; they also involve attacks claimed by terrorist groups, and unclaimed or unresolved attacks. Similarly, The Mother Jones database contains all mass shootings whose overwhelming majority involve a single perpetrator; however, not all shootings by  a single person are categorized as LA terrorism. The following filter is used to distinguish LAs from other terrorist attacks in the GTD and other mass shootings in the Mother Jones database.

\begin{itemize}
    \item For the GTD, the attacks should be planned and committed by a single person or two-person cells (dyads). These two-person cells will be evaluated separately for each attacker since they have been observed to exhibit different behaviors and have been provoked by different environmental responses even for the same incident.
    \item In the GTD, all unclaimed and unresolved attacks are excluded, since these attacks do not provide any data for the attacker's characteristics. 
    \item For the Mother Jones database, the motivations of attackers are checked, and violence stemming solely from a personal grievance are excluded. However, for the sake of further research, we should note that bullying-related attacks have recently been discussed for inclusion as a part of single-issue events~\cite{silva2019fame}. 
\end{itemize}

In the GTD, 70 of the 196 attacks are unclaimed or unresolved, and 84 of the attacks are identified as LA attacks.  In the Mother Jones Database, 17 out of 44 mass shootings satisfy the criteria of being committed by a single person and stemming from an ideology-based grievance. 9 of those mass shootings match with the records of the GTD. In total, data for 190 LAs have been obtained using these three databases.

Using the National Criminal Justice Reference System database as a template, we have gathered data on the following characteristics of the attackers:

\begin{itemize}
    \item \textit{Demographic and socio-economic data:} age, race, gender, marital status, mental health history, employment status and military history.
    \item \textit{Distal characteristics:} criminal history, relation to radical groups, means of radicalization.
    \item \textit{Proximal warning behaviors:} triggering event and leakage.
    \item \textit{Attack decisions:} Target and weapon selection
    \item \textit{Attack consequences:} fatalities and after-attack behavior.
\end{itemize}

The distinction between ``distal characteristics'' and ``proximal warning behaviors'' in~\cite{meloy2016lone,meloy2017trap} is that the distal characteristics belong to the history of an LA before the attack idea and preparations. ``Proximal warning behaviors'' start with ``pathway to violence'' provoked by an ideological and personal grievance. It can be observed that the distal characteristics or proximal warning behaviors we employ do not match one-to-one  to~\cite{meloy2017trap}, since our data do not specify  the warning behaviors such as ``last resort'', ``energy burst'' or ``directly communicated threat''. Using the available data, the distal characteristics are selected in a way to maximize their relevance to radicalization and violence. 

Our data involves four female LAs (two in the pre-9/11 era, two in the post-9/11 era); which is a very small number among all LAs. Hence, any gender-based analysis on LAs would lack sufficient data. Three pre-9/11-era LAs have chosen terrorism as a career and committed multiple attacks over a decade. Excluding these three LAs, the average age of attackers is 35.11 with a standard deviation of 13.73 and a median of 31. The ages range from 15 to 88. Observations from the data indicate that sudden changes in marital status, mental health history, employment status, and military history serve as a triggering event. In fact, the three most common personal triggering events are separation from partner, losing job or student status, and emergence of mental or physical health issues. While we acknowledge that these factors contribute to radicalism and violence, socio-economic and demographic characteristics may be related to an attack in a proximal or distal way. Hence, we embed some of the socio-economic or demographic data into distal and proximal characteristics. However, we try to avoid using demographic and socio-economic data directly in our behavioral characterization for the sake of fairness.

Through the resources~\cite{brynielsson2013harvesting, bakker2011preventing, meloy2011role, phillips2011lone, gill2014bombing, tierney2017spotting, maccauley2008mechanisms,spaaij2010enigma,spaaij2011understanding}, we construct the following exhaustive list of behaviors by breaking down the rest of the data into 25 binary exhaustive characteristics.

\begin{itemize}

\item Criminal history before the attack

\begin{outerenum}[series=myouterlist]
\item No criminal history
\item One offense 
\item Multiple offenses
\end{outerenum}

\item Knowledge of weapons

\begin{outerenum}[resume=myouterlist]
\item Had formal weaponry training
\end{outerenum}

\item Relation to radical groups

\begin{outerenum}[resume=myouterlist]
\item No prior relations with any extremist groups
\item Has contacts with an extremist or a terrorist group/people
\end{outerenum}

\item Means of radicalization

\begin{outerenum}[resume=myouterlist]
\item Self-radicalized
\item Was not self radicalized
\end{outerenum}

\item Triggering event

\begin{outerenum}[resume=myouterlist]
\item A triggering event caused the attack idea
\item No particular triggering event, radicalized incrementally by the socio-political atmosphere
\end{outerenum}

\item Leakage

\begin{outerenum}[resume=myouterlist]
\item No leakage was made
\item Leakage was made offline
\item Leakage was made online
\end{outerenum}

\item Targets

\begin{outerenum}[resume=myouterlist]
\item Civilians
\item Person symbolizing the enemy ideology (politician, religious leader, abortion doctor, etc.)
\item Law enforcement (military, police, etc.) or government officials
\item No targets aimed / symbolic attack
\end{outerenum}

\item Means of attack

\begin{outerenum}[resume=myouterlist]
\item Firearms
\item Other weapons
\end{outerenum}

\item Fatalities

\begin{outerenum}[resume=myouterlist]
\item No fatalities or injuries
\item Only injuries but no fatalities
\item At least one fatality
\end{outerenum}

\item After-attack behavior

\begin{outerenum}[resume=myouterlist]
\item Was able to escape the crime scene
\item Surrendered / was arrested at the crime scene
\item Committed suicide / was killed at the crime scene
\end{outerenum}

\end{itemize}

As aforementioned, the existing literature handles the term ``behavior'' differently. In our study, the term ``behavior'' is defined as the observable actions of the LA~\cite{eshgi2017mathematical}. Hence, to provide semantic clarity, we first introduce the following temporal terminology on attacker behavior: 

\textit{Early behavior and characteristics:} These behaviors and characteristics do not affect the attack directly; they are behaviors or responses before the idea of an attack emerges. The radicalization process is involved in this section. Examples are prior criminal history, childhood abuse, employment status. In some cases, a triggering event is included in this phase which is a mostly personal grievance that results with the attack idea. In other cases, incremental radicalization causes the attack idea without requiring any triggering incidents. Distal characteristics in the TRAP 18 framework~\cite{meloy2017trap} are among the early behavior and characteristics.

\textit{Preparatory and precursor behavior and characteristics: }These types of behaviors and characteristics are the activities or responses that comes with the attack idea such as acquiring or transporting weaponry, leaking intent, capability testing through dry-runs, meetings and other communications, committing fraud to travel to attack locations, etc. Proximal behaviors in the TRAP 18 framework~\cite{meloy2017trap} are among the preparatory and precursor behavior and characteristics.

\textit{Incident-scene behavior and characteristics:} These types of behaviors and characteristics are the activities at the incident scene prior to the attack. Examples include following an abnormal trajectory, counter-surveillance related or cycling behavior, wearing suspicious clothing (a trench coat on a 95-degree day, sunglasses on a rainy day, etc.). Incident-scene behavior is analyzed under abnormal or unusual trajectory. Inevitably, information at this detail is not available publicly or in literature, since attacks are either limitedly or partially caught on grainy security footage or not caught at all. 

\textit{After-attack behavior and characteristics:} These types of behaviors and characteristicss involve the behavior of the terrorist right after the attack. Some examples include escaping the scene, committing suicide.
The available literature mostly focuses on early, preparatory and after-attack behavior~\cite{brynielsson2013harvesting, bakker2011preventing}. 

The first 8 characteristics in the list involve the attacker's \textit{early behavior and characteristics}. In this period, having prior criminal records and formal weaponry training are factors that are found to increase the number of casualties~\cite{ellis2016analysing}. In fact, Capellan et al.~\cite{capellan2015lone} emphasized that some LAs enroll in the formal weaponry training units just to enhance their attack capabilities. The locus of radicalization also differs among attackers. An LA, who is responsible for 16 bank robberies and two bombings, was radicalized by his parents since childhood, while another one was  a member of Al-Qaeda. Therefore, while some LAs have prior contacts to extremist or terrorist organizations; others have no connections to radical organizations but to radical people. Furthermore, some LAs do not have connections to even radical people, they are radicalized by their own ``gaslighting''. For example, another LA opened fire to Family Research Council headquarters in 2013, initiated his radicalization process by himself through biased research from various websites. The means of radicalization characteristic distinguishes LAs as being initiated to radicalization by other people or by their own effort. 

The attack decision may or may not stem from a triggering event. The triggering event can be personal such as being fired from a job, or social events such as 9/11-attacks~\cite{hammandspaaij}. While some LAs do not require a triggering event but are radicalized by incremental and cumulative life experiences, some others require being triggered multiple times in order to develop the attack intent. In our database, 99 out of 152 LAs have certain triggering events initiating the attack idea. Out of these 99 LAs, 67 of them were triggered by a personal event. Two of these LAs were triggered by multiple personal events. 30 LAs were triggered by a social event. Two LAs were triggered both by a social and a personal event. Personal trigger events are broken down as in Table \ref{tab:table1}.   

\begin{table}[h]
\begin{tabular}{|l|c|}
\hline
\multicolumn{1}{|c|}{\textbf{Personal Event}}    & \textbf{Frequency} \\ \hline
Separating from partner/wife/family              & 13                 \\ \hline
Losing job or dropping out of school             & 10                 \\ \hline
Mental/physical health problems                  & 8                  \\ \hline
Arrest-related issues                            & 6                  \\ \hline
News / poster / graffiti / newspaper irritation  & 5                  \\ \hline
Sting operations                                 & 5                  \\ \hline
Rejection by friends / colleagues / other people & 4                  \\ \hline
Financial or social security problems            & 3                  \\ \hline
Eviction / homelessness                          & 3                  \\ \hline
Fight with neighbors                             & 3                  \\ \hline
Denied applications                              & 2                  \\ \hline
Deployment as an army member                     & 2                  \\ \hline
Travel                                           & 2                  \\ \hline
Online discussions                               & 1                  \\ \hline
Unreturned calls                                 & 1                  \\ \hline
Gaslighted by partner                            & 1                  \\ \hline
\end{tabular}
\caption {Personal Trigger Events and Their Frequencies}
\label {tab:table1}
\end{table}

 Leakage is also a common trait of attackers; many LAs leak their intent before the actual attack. One extreme case belongs to an LA who posted a 1500-page manifestation on social media the night before the attack~\cite{brynielsson2013harvesting}. These leaks can be offline through chats, letters, etc.; or online through social media, e-mails, etc. Our definition of online leakage involves the cases where the LA benefits from the Internet to broadcast attack intent; hence, phone calls or television broadcasts are considered as offline. 

Target selection is another attack characteristic and the targets are civilians in most attacks. However, if the LA fixates on a person that symbolizes the enemy ideology; the attacker may choose to spare other civilians. The target can also be the security forces, military, or a formal government official on duty~\cite{becker2014explaining}. Most LA attacks are conducted with firearms which are shown to be more deadly~\cite{ellis2016analysing}. Other weapons include explosives, blades, bodily weapons (hand, feet, etc.), or vehicles (trucks, cars, etc.). 

\subsection{LA Classification}\label{classes}


Using 9/11  as a cutoff point for understanding recent exacerbation in terrorism, we first compare pre-9/11 and post-9/11 LAs to detect the changes in LA terrorism over time. Then, focusing on post-9/11 LAs,  we classify them in multiple domains. The first classification domain is offered by Gill et al.~\cite{gill2014bombing}, that is, the ideological classification: jihadists, right-wing LAs, and single-issue LAs.

The second classification domain is clustering according to incident-scene behavior. However, incident-scene visuals or data are not publicly available. Serious game design simulating real-world conditions is a suitable surrogate and a widely-used approach for the resolution of such predicaments~\cite{roungas2018future}. In the absence of available data, at the Game Research for Information SecuriTy (GRIST) Lab~\footnote{http://gursoy.rutgers.edu/GRIST/index.html} at Rutgers, we have developed a 2D-game where players can imitate the trajectory and target selection of an attacker or the strategy of a defender aiming to catch the attacker.  The game considers the effect of human dynamics and crowd flow on target selection. In this game, the attacker moves over a network and knows the density of each adjacent node before he makes his selection to move to an adjacent node or stay at the same node, or attack (see Figure \ref{fig:game}). On the other hand, the defender patrols the network without knowing the exact location of the attacker. If both players are at the same node, the defender detects the attacker with some probability. Data from playthroughs are collected for each session. 

\begin{figure}[H]
	\includegraphics[scale=0.5]{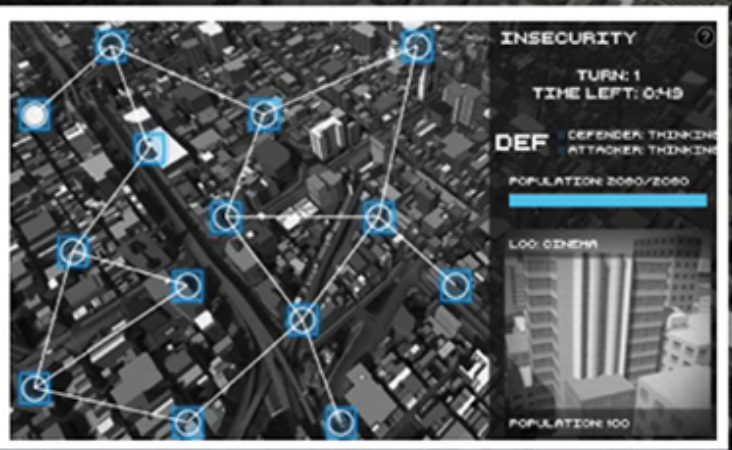}
	\caption{Playable game developed in the GRIST Lab.}
	\label{fig:game}
\end{figure}

\noindent 101 game sessions (15 two-people runs, 69 against greedy-AI and 17 against improved AI) are analyzed using simple clustering tools (hierarchical clustering) and results are compared in order to select the best features for clustering. In terms of the attacker, the important classifiers are determined as: i) time of planting the bomb, ii) distance between the attacker and the defender, iii) node's occupancy rank, and iv) node's centrality. In the clustering results, the occupancy rank of nodes has emerged as more important than the actual population itself. According to these features, 5 types of attackers are extracted (Table \ref{tab:figure2}): i) maximum damagers, ii) symbolic attackers, iii) daredevils, iv) attention seekers, and iv) stallers.

\begin{table}[h!]
	\begin{tabular}{c}
		\includegraphics[scale=0.2]{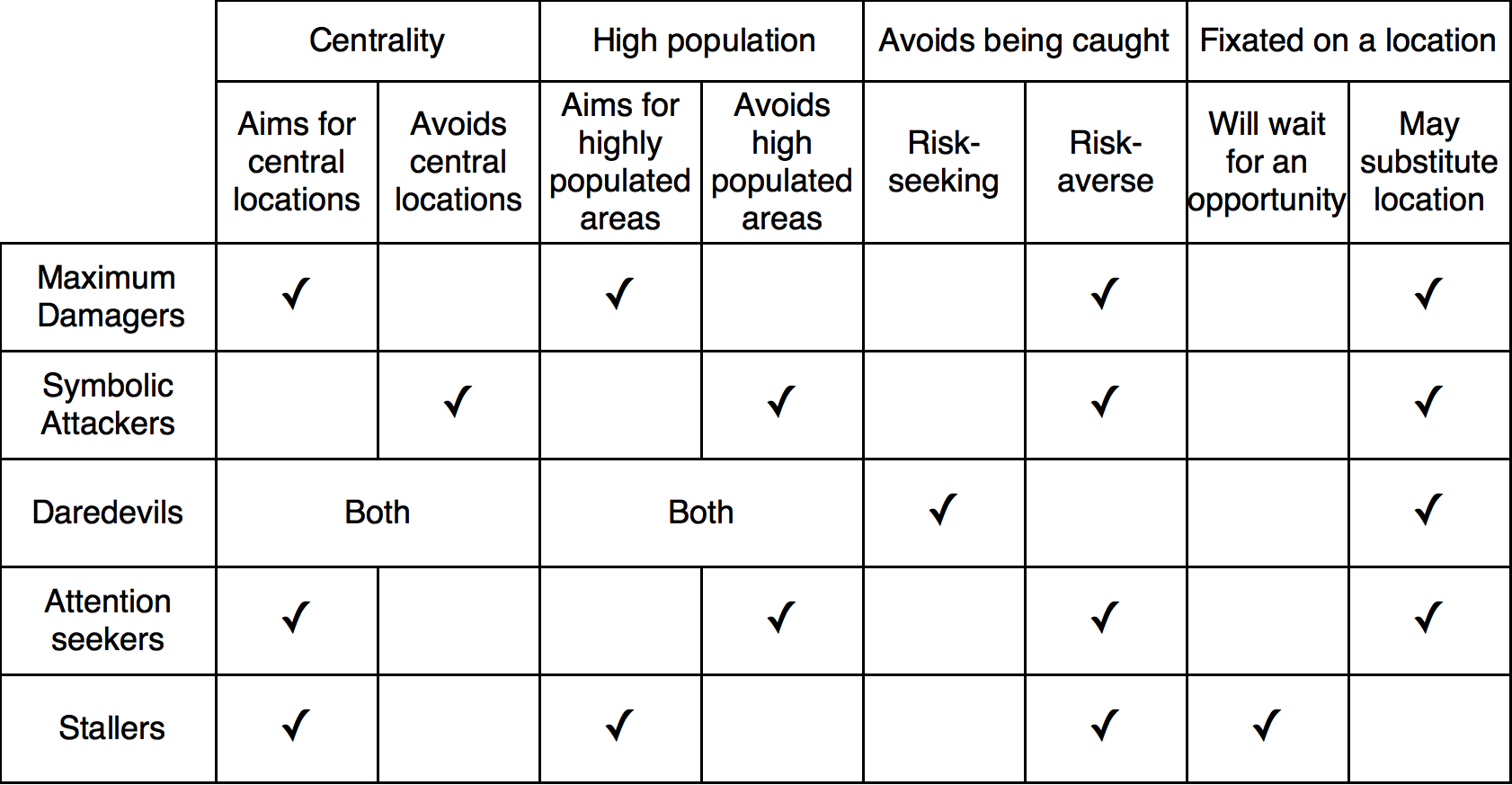}
	\end{tabular}
	\caption{LA-types using incident-scene behavior obtained from the playable game.}
	\label{tab:figure2}
\end{table}

The maximum damagers follow the nodes to reach a node with the highest population, whereas stallers tend to keep a spiral route and wait for an opportunity to attack a fixated target. Daredevils stay close to the defenders in a risk-seeking manner.

The real-world counterparts of these LA types are derived from the correlations between game and real-world data. Real-world maximum damagers aim to harm as many people as possible or conduct a series of attacks. Real world symbolic attackers target a person or a leader that symbolizes the ``enemy'' ideology or send a message without causing any harm by attacking fake weapons. Daredevils attack directly security forces where being unsuccessful is highly likely or go on a killing spree without a plan; whereas, attention seekers attack central locations at odd hours without intending heavy casualties. Finally, stallers fixate on a location or a person, stalk the target, and plan the attack. 

Finally, the third classification domain is in terms of behaviors. In order to obtain behavioral classes, 152 post-9/11 LAs are clustered by a K-Means algorithm. The best results are obtained with 7 clusters with distinct characteristics. 3 of these clusters contain a small and insufficient number of LAs, and hence, we will operate on the 4 clusters involving sufficient number of LAs.

\section{Methodology}\label{methodology}

\subsection{A-priori Algorithm}\label{apriori}

The \textit{A-priori} algorithm was proposed by Agrawal and Srikant~\cite{agrawal1994fast}, and has been successfully applied to many social problems. Nijkemp et al.~\cite{nijkamp2008economic} implemented it to identify the association rules for the valuation  of different biodiversity indicators and different kinds of habitats, while Parack et al.~\cite{parack2012application} analyzed the relationship between grading system and attendance in an educational setting. A security-based application of the \textit{A-priori} algorithm is ~\cite{nazeri2001experiences}, which analyzed the relationship between different classes of passengers and safety factors.

 In this study, the \textit{A-priori} algorithm is applied to extract the associations among behavioral and attack characteristics. For LAs, an association rule of the form $A \rightarrow B$ can be interpreted as \textit{``If behavior $A$ exists, then behavior $B$ also exists''}. A rule chain is of the form $A \rightarrow B \rightarrow C$, meaning that \textit{``If behavior $A$ exists, then behavior $B$ also exists. If Behavior $B$ exists, then behavior $C$ also exists.''}. It should be noted that association rules do not imply a cause-effect relationship; these rules state that given one behavior, another behavior may also exists. Mathematical derivation of association rules are explained and every mathematical formula is presented in Appendix~\ref{appendix:apriori}.

Let $\mathcal{S}$ be a set of $d$ items, i.e.  $\mathcal{S}=\{s_1, s_2, \dots, s_d\}$, and $\mathcal{T}$ be a set of $n$ transactions in a database, i.e. $\mathcal{T}=\{t_1, t_2, \dots, t_n\}$. A transaction involves at least one item. In the LA terrorism setting, the transactions are LAs, and the items are one of the 25 binary characteristics. As we have constructed the list of characteristics in an exhaustive manner, a given LA satisfies at least one condition on the list. In our case, $d=25$ representing each one of the 25 binary characteristics, and $n$ is 190 for overall LA evaluations, 38 for pre-9/11 LAs, 152 for post-9/11 LAs since it is the number of LAs that are available in each category.

In order to establish an association rule  of the form $A \rightarrow B$ (denoting that a characteristic implies another), two item-sets must employ three thresholds to pass: {\it support, confidence, and lift.\/} The support of $A$ is defined as the fraction of all transactions involving $A$, and the support of $B$ is similarly defined. In the LA setting, the {\it support\/} threshold checks if these characteristics, are one-by-one frequent enough so that we can generalize these characteristics. If the {\it support\/} exceeds a predetermined threshold, then the itemsets are assumed to be frequent enough. 

If $A$ and $B$ are frequent enough, then the test for $A \rightarrow B$ is subjected to a confidence test. {\it Confidence\/} is the fraction of the transactions where $A$ and $B$ are together in the transactions that involve $B$. In the LA setting, the {\it confidence\/} is the fraction of number of LAs who also have the characteristic $A$ among LAs who have the characteristic $B$. If the {\it confidence\/} exceeds a predetermined threshold, then coincidentally or not, $A$ implies $B$. 

Another measure, {\it lift\/}, which is also called the {\it interestingness\/} factor, controls if this implication is coincidental. It compares the number of transactions that $A$ and $B$ are together to the multiplication of the number of transactions that $A$ and $B$ exist  marginally. If the number of transactions that $A$ and $B$ are together is higher, then the implication is not coincidental, and $A \rightarrow B$.

Another {\it interestingness\/} measure is \textit{cohesion} which is a substitute for lift. It uses the entropy concept that measures the disorder or uncertainty in data. While lift controls if $A$ and $B$ together frequent enough for the rule $A \rightarrow B$, entropy controls whether $A$ is not frequent enough in the absence of $B$. Then, cohesion is calculated as an inverse measure of entropy. If the cohesion exceeds a predetermined threshold, it can be concluded that $A$ implies $B$. We refer the readers to Appendix~\ref{appendix:signi} for mathematical details.

\subsection{Parameter Selection}\label{pselect}

We define a \textit{strong association rule} or a \textit{strong implication} as rules with a minimum support threshold of 20\% and a minimum confidence threshold of 70\%. A \textit{two-way association rule} or a \textit {two-way implication} is the case where $X \rightarrow Y$ and $Y \rightarrow X$, and it is denoted by $ X \leftrightarrow Y$. A  \textit{strong two-way association rule} or a \textit {strong two-way implication} where both $X \rightarrow Y$ and $Y \rightarrow X$ exceed a minimum support threshold of 20\% and a minimum confidence threshold of 70\%. It should be noted that the literature on \textit{A-priori} association rules mostly implement a minimum support threshold level of 10-15\% ~\cite{sarath2013association,danping2011data}. However, LA events are rare, making LA data a small dataset. 10-20\% of the data refers to a number that is insufficient for generalization. Hence, the minimum support threshold is increased to 50\%. If the support threshold is too low, the emerging characters will be in small numbers that will prevent generalization. If it is too high, then some associations will be lost by dropping important characters due to a high filter rate. The minimum support and minimum confidence threshold values are determined by tuning in a way that provides both sufficient number of data and emergence of important association rules.

The \textit{A-priori} algorithm parameters (support and confidence) aim to balance generating non-significant association rules (type 1 error) and not missing significant ones (type 2 error) ~\cite{hamalainen2008efficient}. In our case, the minimum confidence threshold is 0.7 which is a relatively low value. Such a confidence level avoids missing significant rules, at the expense of producing non-significant ones. Once a rule is constructed in the form of $A \rightarrow B$, it requires a further statistical test to determine the significance of a rule. This test has the null hypothesis that $A$ and $B$ are independent. In order to check independence, the number of occurences of $A$ and $B$ are compared to the expected number of occurences when they are independent. If the null hypothesis is rejected, then a proof for the associations have been found. We refer the readers to Appendix~\ref{appendix:signi} for mathematical details. 

\subsection{Chain Rules (R-Rules)}\label{rruless}

An R-rule, also known as a rule of rules, is a hyper-rule in the form  $(A \rightarrow B) \rightarrow (C \rightarrow D)$. An association chain (implication chain) is a special type of R-rules in the form $(A \rightarrow B) \rightarrow (B \rightarrow \mathcal{...}) \rightarrow (\mathcal{...} \rightarrow C)$ or $A \rightarrow B \rightarrow ... \rightarrow C$. In other words, an association chain is a statistically significant aggregation of association rules in the form of $A \rightarrow B \rightarrow ... \rightarrow C$. 

The statistical significance of an R-rule is measured by cohesion~\cite{gras2006discovering}. While the item-wise cohesion measure checks whether $A$ does not occur frequently enough without $B$, the rule-wise cohesion measure uses the cohesion for if $A \rightarrow B$, $B \rightarrow C$, and $A \rightarrow C$ holds. We refer the readers to Appendix~\ref{appendix:rrules1} for mathematical details.

\section{Association Rules for LA Behavior} \label{application}

The following subsections present the outputs of the \textit{A-priori} algorithm results. In Section \ref{section:overall}, we provide the \textit{A-priori} algorithm results for all 190 LAs in the database, and Section \ref{section:temporals} compare pre- and post-9/11 LAs, and displays the temporal change in LA characteristics. Bakker and de Graaf~\cite{bakker2011preventing} conclude that LAs yield more common characteristics when analyzed by their ideologies. Referring to this study, in Section \ref{section:motivate}, we analyze LAs according to their ideological motivations. We also compare these results to incident-scene-based classification (Section \ref{section:incidentscene}) and behavior-based classification (Section \ref{section:behavioralc}).   

\subsection{Overall evaluations} \label{overall}

Analyzing 190 LAs over 60 years, we have found the following  most common characteristics:

\begin{itemize}
\item 62.1\% of all LAs required a triggering event,
\item 62.1\% of all LAs targeted civilians,
\item 59.5\% of all LAs committed their attacks using firearms,
\item 52.1\% of LAs had no prior connections to extremist/terrorist people,
\item 51.0\% of LA attacks were fatal.
\end{itemize}

The overall analysis of 190 LAs produce few common characteristics and none them belong to the early behavior stage. Hence, these common characteristics are not viable in capturing the early signs of an LA. Figure \ref{fig:figure3} displays the associations among common characteristics found by the \textit{A-priori} algorithm. These results are: 
\begin{itemize}
\item One strong two-way association emerges: the usage of firearms imply fatalities, and vice versa. 
\item  Only two strong one-way associations are available for all LAs.
\begin{itemize}
\item Usage of firearms implies a triggering event. 
\item Fatalities imply civilian targets. 
\end{itemize}
\end{itemize}

Even though some commonalities can be identified, the face of LAs have changed significantly through time. To capture these changes, we will first compare pre-9/11 LAs to post-9/11 LAs.

\begin{figure}[H]
	\includegraphics[scale=0.2]{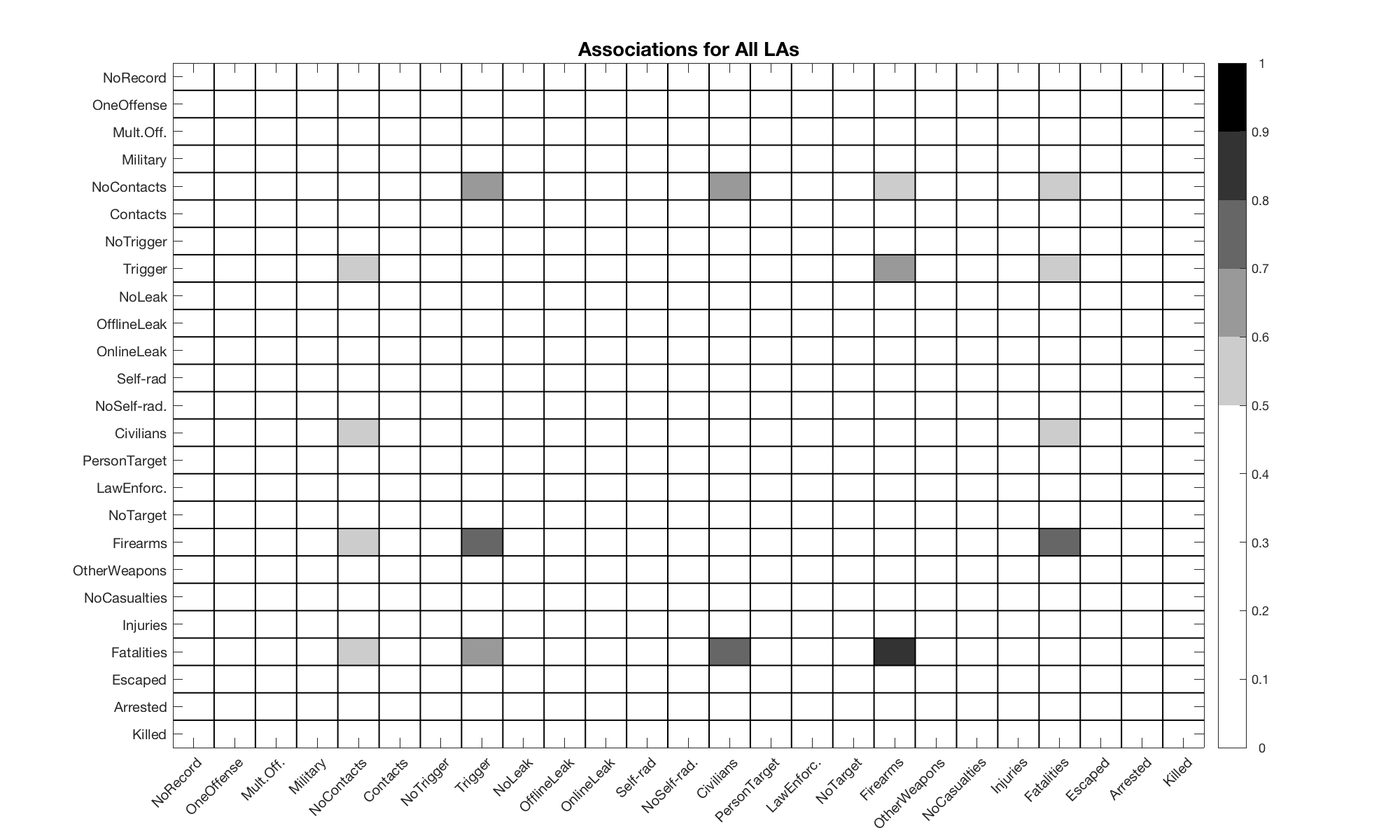}
	\caption{Associations for all LAs. For this figure and all forthcoming figures for associations, the color of each cell indicates the confidence level of the association. The color white indicates three possibilities: $i)$ the confidence of the rule is less than 0.5, $ii)$ the confidence of the rule is greater than 0.5 but the lift is less than 1, therefore, the associations are coincidental, and $iii)$ at least one item is not frequent enough to construct an association. The shades of gray indicate the magnitude of the confidence, where darker shades specify a stronger association.}
	\label{fig:figure3}
\end{figure}

\subsection{Comparison of Pre-9/11 and Post-9/11 LAs} \label{temporals}

Analyzing all LAs has yielded sparse relationships as shown in the previous section. One reason for this sparsity is that the triggers, technologies, opportunities, and other conditions that lead an LA to an attack has changed significantly over the years. Hence, in this subsection, we present the associations for pre-9/11 and post-9/11 LAs in a way to show how the LA terrorism has changed. The database holds 38 LAs before 9/11 and 152 LAs after 9/11 and the \textit{A-priori} algorithm was separately applied to these two data sets. We present the figure of associations for both era in Figure \ref{fig:temporal2}. The density of gray-shaded cells indicate that pre-9/11 LAs had more common characteristics than post-9/11 LAs. In the following subsections, we delve deeper into both time periods and present the results of the \textit{A-priori} algorithm results.

\begin{figure}[H]
    \centering
     \begin{subfigure}[b]{\textwidth}
         \centering
         \includegraphics[width=\textwidth]{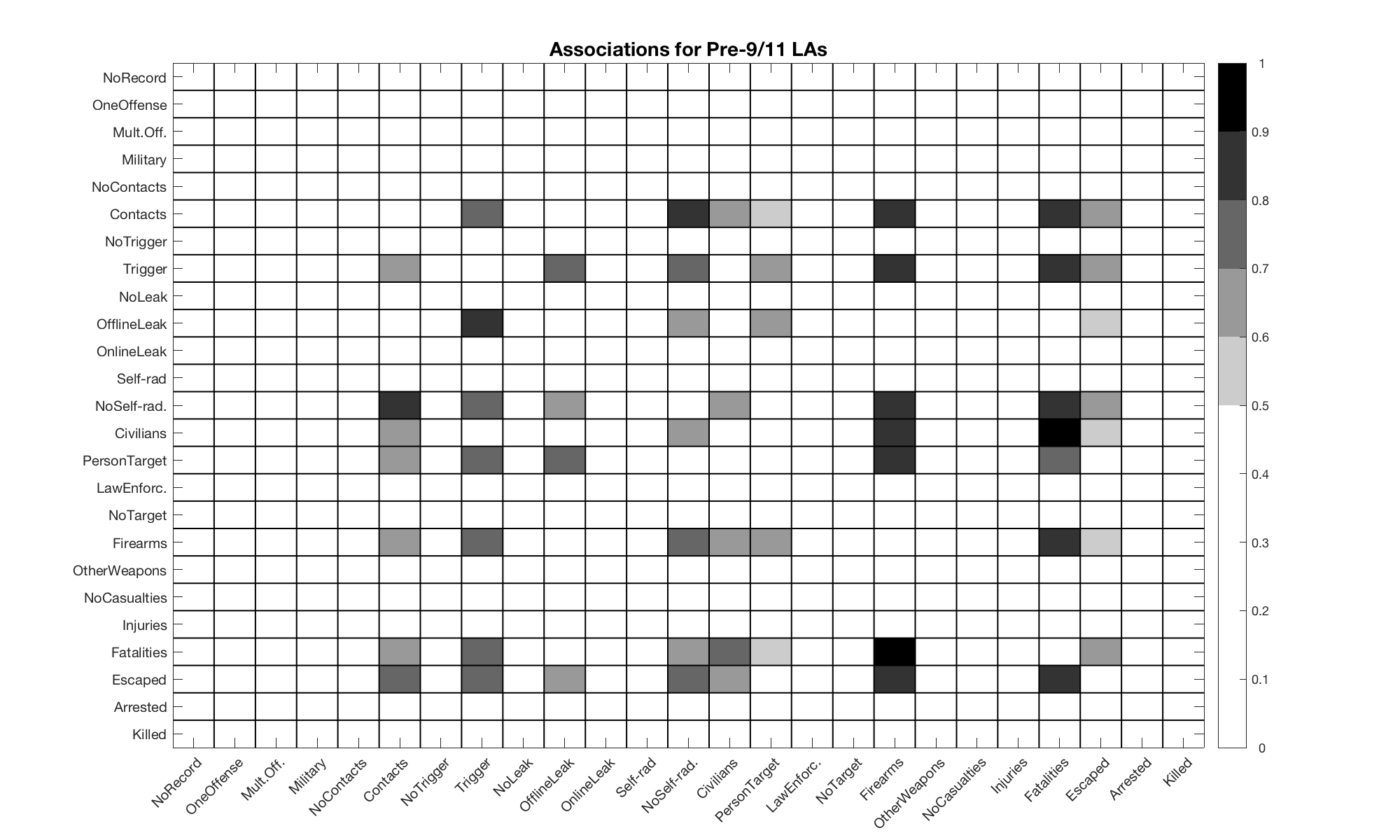}
         \caption{Pre-9/11 LAs}
         \label{fig:pre911}
     \end{subfigure}
    \hfill
     \begin{subfigure}[b]{\textwidth}
         \centering
         \includegraphics[width=\textwidth]{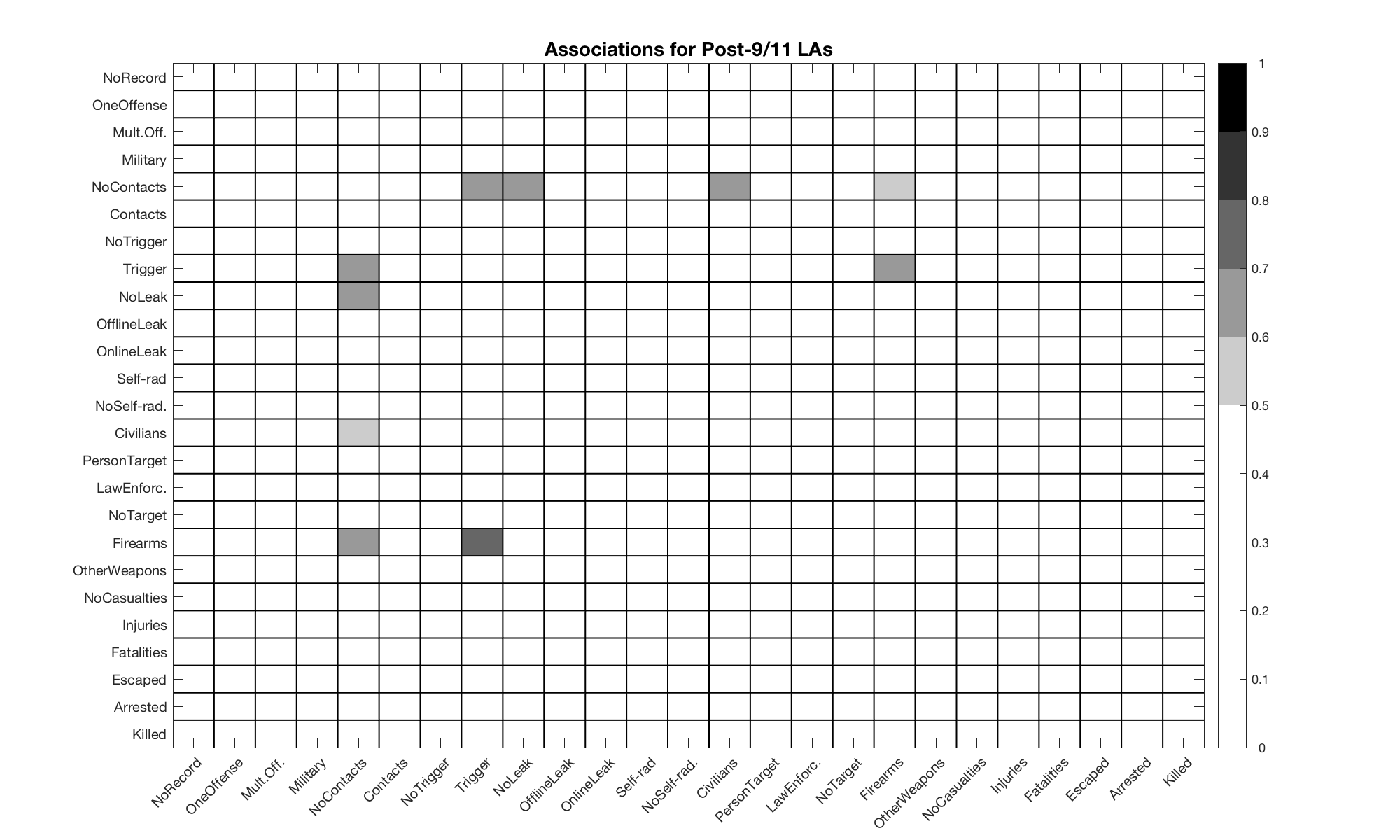}
         \caption{Post-9/11 LAs}
         \label{fig:post911}
     \end{subfigure}
        \caption{Associations denoting temporal change in LA behavior. Pre-9/11 LAs having more associations than post-9/11 LAs shows that the LAs have diversified greatly in terms of behavior and characteristics.}
        \label{fig:temporal2}
\end{figure}

\subsubsection{Associations in the pre-9/11 era} 

We have found the following  most common characteristics
\begin{itemize}
\item 81.6\% of the pre-9/11 LAs used firearms in their attacks.
\item 76.3\% of the pre-9/11 LA attacks were fatal.
\item 68.3\% of the pre-9/11 LAs had a triggering event that led to attack idea.
\item 65.8\% of the LAs leaked intent offline.
\item 65.8\% of the LAs were not self-radicalized.
\item 63.2\% of the LAs had prior contacts with extremist or terrorist groups.
\end{itemize}

\noindent The algorithm produces eight strong two-way associations:
\begin{itemize}
\item The existence of a triggering event implies the usage of firearms, and vice versa.
\item Having prior contacts to extremist/terrorist groups implies not being self-radicalized, and vice versa.
\item  The existence of a triggering event implies offline leakage, and vice versa. 
\item Not being self-radicalized implies a trigger event, and vice versa.
\item The existence of a triggering event implies offline leakage, and vice versa.
\item Not being self-radicalized implies usage of firearms, and vice versa.
\item Civilian targets imply fatalities, and vice versa.
\item Usage of firearms implies fatalities, and vice versa.
\end{itemize}

\noindent Strong one-way associations are: 
\begin{itemize}
\item Having prior contacts to extremist/terrorist groups are implied by being able to escape from the crime scene. 
\item The existence of a triggering event and the usage of firearms are implied by any of the following: having prior contacts to extremist/terrorist groups, offline leakage, targeting a person who symbolizes an enemy ideology, being able to escape from the crime scene. 
\item Offline leakage is implied by targeting a person who symbolizes an enemy ideology.
\item Fatalities are implied by having prior contacts to extremist/terrorist groups, offline leakage, targeting civilians or a person who symbolizes an enemy ideology, being able to escape from the crime scene.
\end{itemize}

\subsubsection{Associations in the post-9/11 era}

In this era, the LA attack types and \textit{modi operandi} have diverged, and as a result there is only one common characteristic 
\begin{itemize}
    \item 60.5\% of the LAs had a triggering event that led to the attack idea.
\end{itemize}

\noindent Interestingly, no strong two-way associations are obtained for this time period and as can be seen from Figure \ref{fig:post911}, only one strong one-way association exists for post-9/11 LAs:

\begin{itemize}
    \item The usage of firearms implies a triggering event.
\end{itemize}
\textbf{Comparison of Pre-9/11 and Post-9/11 LAs}.

The common characteristics and the associations show that pre-9/11 LAs had more connections to extremist/terrorist people or groups than post-9/11 LAs. This indicates that while violence once was a means chosen by extremist-adjacent people, it has now trickled down to emerge as an option even for people without any radical connections. This result also holds for formal weaponry training (military or otherwise). While pre-9/11 LAs had a rate of 47\% in formal weaponry training, this characteristic applies to 22\% of post-9/11 LAs, indicating that terrorism has spread into the realm of ordinary civilians more and more. Furthermore, LA terrorism has become more diverse in preparatory behaviors. While both pre- and post-9/11 LAs target civilians more, the rate of other groups has significantly risen. Targeting the law enforcement and government officials has increased from 13\% to 29\%. Similarly, weapon selection has also changed significantly. The usage of firearms has dropped from 89\% to 53\%. Post-9/11 LAs weapon choices vary from explosives to hatchets, machetes, as well as to public or personal vehicles. The variety in target and weapon selection has led to varieties in the aftermath of the attack. While pre-9/11 attacks had a higher fatal attack rate (76\%), post-9/11 LAs commit less attacks that have fatal consequences (47\%). However, the number of fatalities per attack does not exhibit statistically significant differences. On the contrary, the standard deviation in the number of fatalities has more than doubled for post-9/11 LAs (from 3.60 to 8.43); meaning that, while the number of unsuccessful attacks has increased due to sting operations or improved technology, post-9/11 LAs have caused more mass casualty than pre-9/11 LAs. In terms of after-attack behavior, the rate of escapes significantly decreased while the rate of suicides/killings has increased due to the technological advances. For example, in 2016, the police killed an attacker with a remote-controlled bomb disposal robot, which was the first time that U.S. law enforcement used a robot to subdue a terrorist~\cite{lewis2015human}. 

\subsection{Evaluations regarding attacker motivation} \label{motivate}

According to Bakker and de Graaf~\cite{bakker2011preventing}, classifying LAs into three ideological segments (jihadists, right-wing LAs, single-issue LAs) produces numerous distinct characteristics. In our data, these three groups hold the 88\% of the post-9/11 LAs. Grouping LAs according to their ideology reveals more similarities and stronger implications than overall comparisons. Among the 152 post-9/11 LAs, 40 are jihadist LAs, 58 are right-wing LAs, and 36 are single-issue LAs.

\subsubsection{Association Rules for Jihadist LAs}

Among 40 jihadist LAs, We have found the following most common characteristics 
\begin{itemize}
\item 67.5\% of jihadist LAs do not leak the attack intent.
\item 62.5\% of jihadist LAs choose weapons other than firearms as the means of attack (mostly explosives). 
\item 70\% of jihadist LAs target civilians. Together with the law enforcement targets, they make up 97\% of jihadist LAs' targets. 
\item Half of the jihadist LAs are not self-radicalized and half of them do not need a triggering event for to exhibit attack intent. 
\end{itemize}

\noindent We have found only one strong two-way association:
\begin{itemize}
\item Having prior contacts to extremist/terrorist groups implies not being self-radicalized, and vice versa.
\end{itemize}
\noindent We have also found one-way associations which are summarized as  (Figure \ref{fig:figure6}): 
\begin{itemize}
\item Civilian targets are implied by no fatalities after the attack or no intent leakage.
\item No injuries/fatalities after the attack are implied by being arrested at the crime scene. 
\item Being arrested at the crime scene is implied by not being self-radicalized. 
\end{itemize}

\subsubsection{Association Rules for Right-Wing LAs}

Among 58 post-9/11 right-wing LAs, we have found the following most common characteristics 
\begin{itemize}
\item 84.4\% of the right-wing LAs target civilians.
\item 70.7\% of the right-wing LAs require a triggering event for the attack intent. 
\end {itemize}

\noindent No strong two-way associations emerges among these LAs and we only find the following strong one-way associations which are (Figure \ref{fig:figure7}):
\begin{itemize}
\item Civilian targets are implied by not having prior contacts to extremist/terrorist groups.
\item A triggering event is implied by either not having prior contacts to extremist/terrorist groups or usage of firearms.
\end{itemize}

\subsubsection{Association Rules for Single-Issue LAs}

Among 36 post-9/11 single-issue LAs, we have found the following most common characteristics 
\begin{itemize}
\item 72.2\% of these single-issue LAs have no prior contacts to extremist/terrorist groups or people.
\item 61.1\% of these LAs require a traumatic triggering event to intend the attack.
\end{itemize}

\noindent Only one strong two-way association emerges:
\begin{itemize}
\item Usage of firearms implies law enforcement/government official targets, and vice versa.
\end{itemize}
\noindent Only two strong one-way associations emerge (Figure \ref{fig:figure8}):
\begin{itemize}
\item Not having prior contacts to extremist/terrorist groups is implied by either law enforcement/government official targets or usage of firearms.
\item Law enforcement/government official targets also imply the existence of a trigger event or no intent leakage.
\end{itemize}

\begin{figure}[H]
    \centering
     \begin{subfigure}[b]{\textwidth}
         \centering
         \includegraphics[width=0.65\textwidth]{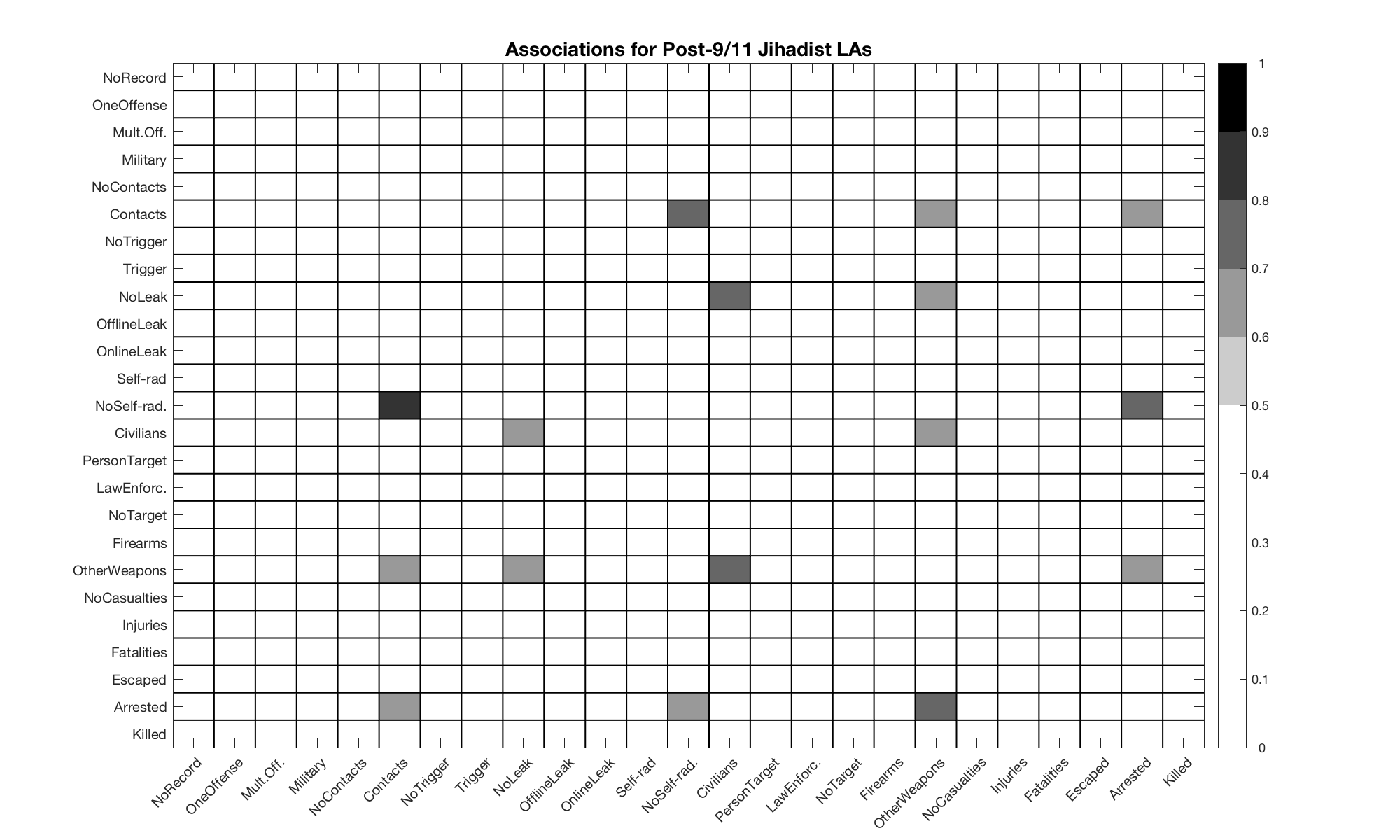}
         \caption{Post-9/11 Jihadist LAs}
         \label{fig:figure6}
     \end{subfigure}
    \hfill
     \begin{subfigure}[b]{\textwidth}
         \centering
         \includegraphics[width=0.65\textwidth]{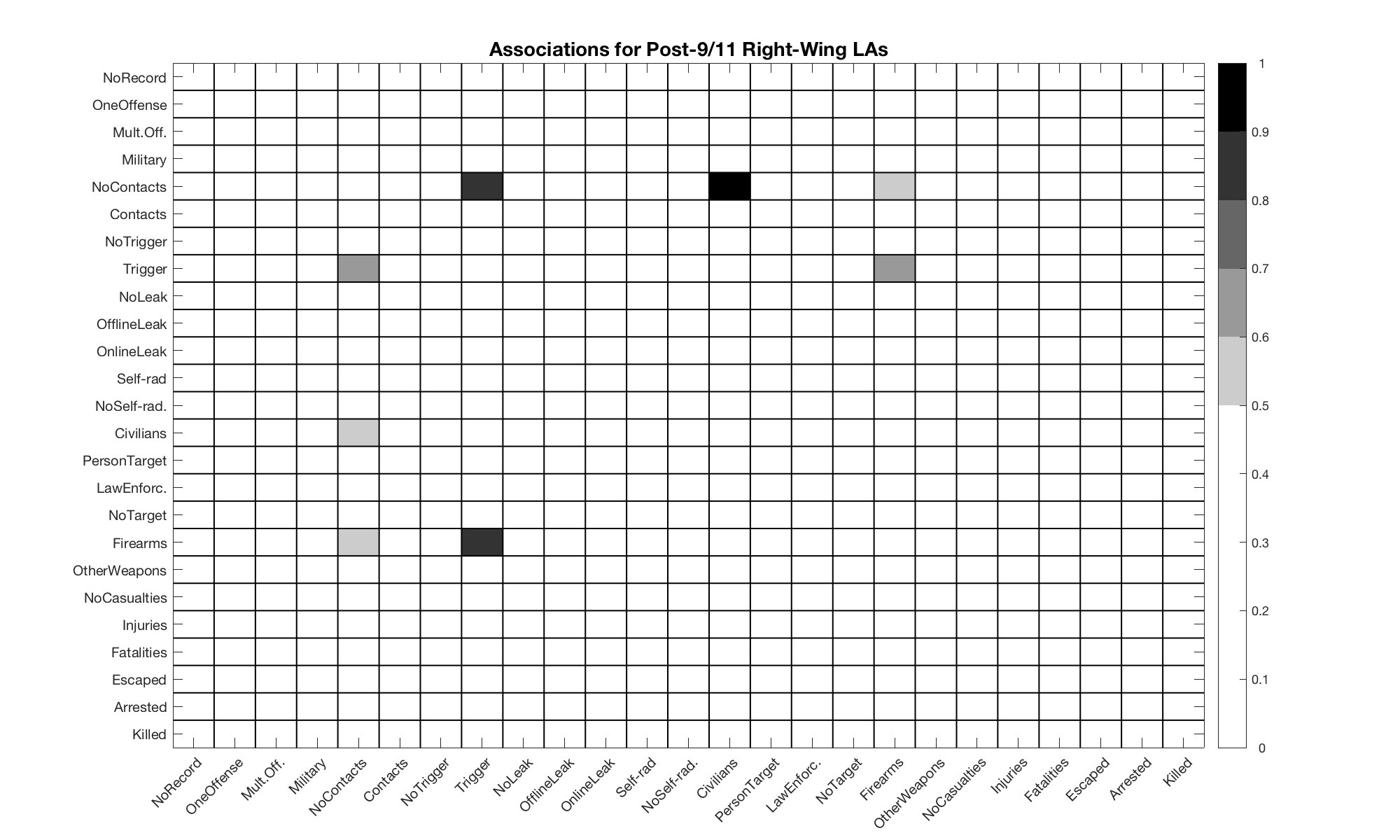}
         \caption{Post-9/11 Right-Wing LAs}
         \label{fig:figure7}
     \end{subfigure}
     \begin{subfigure}[b]{\textwidth}
         \centering
         \includegraphics[width=0.65\textwidth]{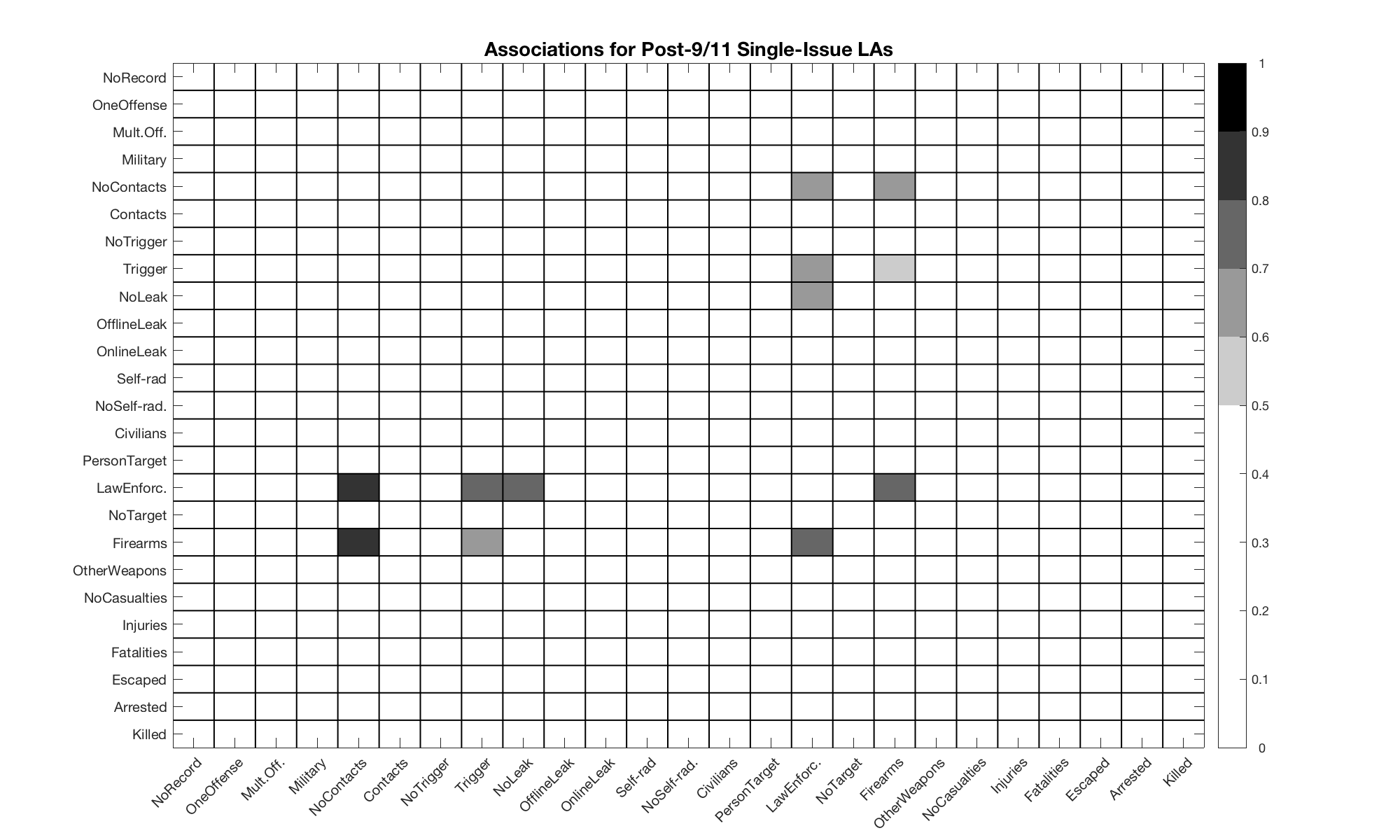}
         \caption{Post-9/11 Single-Issue LAs}
         \label{fig:figure8}
     \end{subfigure}
        \caption{Associations by ideological motivations. Even though each ideological class has distinctive demographic distinctive characters as given in~\cite{bakker2011preventing,gill2014bombing,lansdekilde2019radicalization}, this classification adds very little to the post-9/11 associations and the \textit{A-priori} algorithm does not find many associations.}
        \label{fig:temporal1}
\end{figure}

\subsection{Evaluations regarding incident-scene behavior}\label{incidentscene}

Given that ideological classification has not provided distinct behavioral results, another perspective for behavior analysis becomes necessary. Such a classification also allows us to evaluate the reflection of incident-scene behavior in early, preparation, and after-attack behaviors. Out of 152 post-9/11 LAs, 54 of them are maximum damagers, 37 are symbolic attackers, 27 are daredevils, 27 are attention seekers, and 7 are stallers. The characteristics of each type are presented in Table \ref{tab:figure13}. Since the number of data for stallers are statistically insufficient for analysis, we will focus on the other four groups for associations.

\begin{table}[ht]
\caption{Class comparisons for incident-scene behavior.}
\centering
\begin{tabular}{cc}
\includegraphics[scale=0.2]{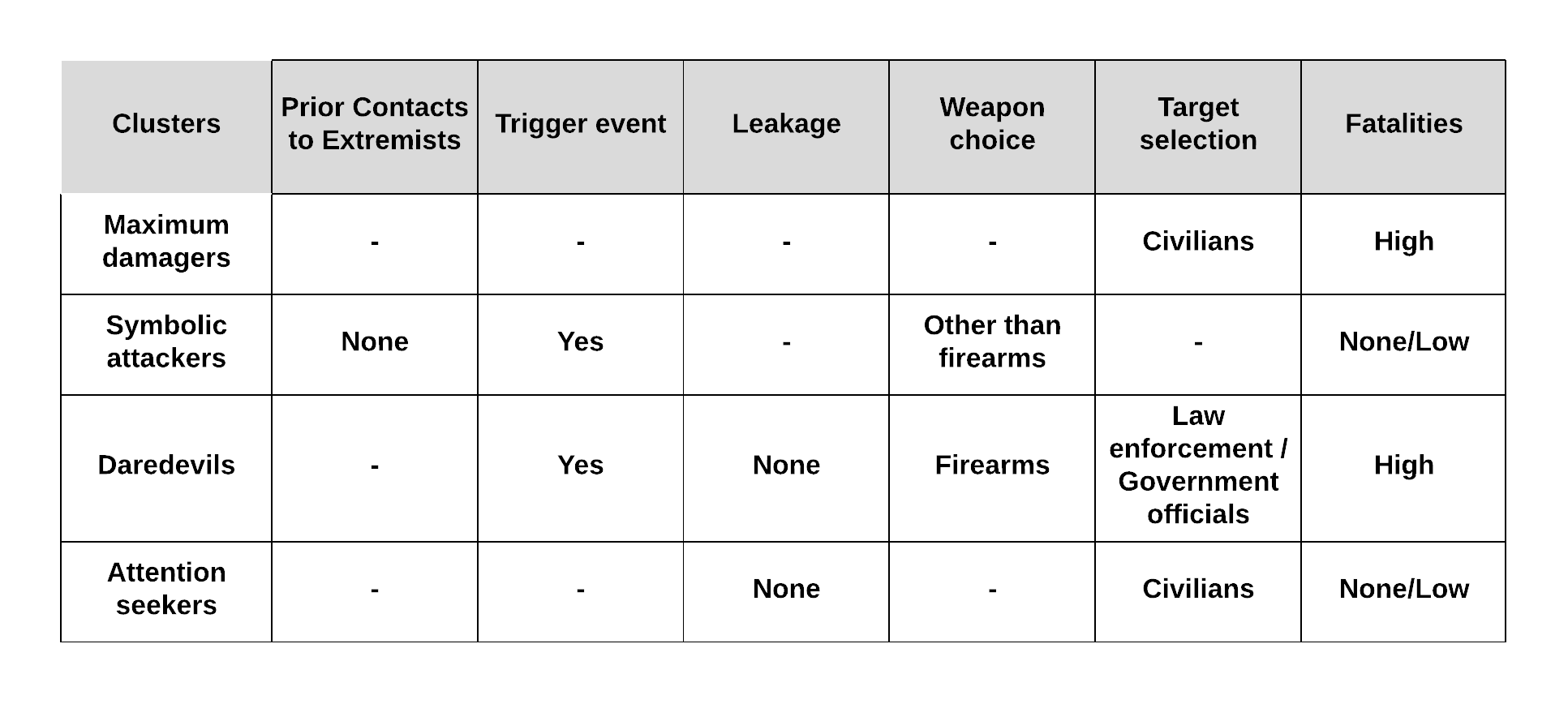}
\end{tabular}
\label{tab:figure13}
\end{table}

\subsubsection{Association Rules for Maximum Damagers}

Among 58 post-9/11 maximum-damager LAs, we have found two common characteristics:
\begin{itemize}
\item 92.5\% of the maximum damagers target civilians.
\item 61.1\% of the maximum damagers attacks are fatal.
\end{itemize}

\noindent One strong two-way associations emerges:
\begin{itemize}
    \item The usage of firearms implies fatalities, and vice versa. 
\end{itemize}
\noindent We have found the following strong one-way associations (Figure \ref{fig:figure9}):
\begin{itemize}
\item Not having prior contacts to extremist/terrorist groups implies civilian targets and fatalities.
\item A triggering event involves fatalities.
\item The usage of other weapons implies civilians.
\end{itemize}

\subsubsection{Association Rules for Symbolic Attackers}

Among 37 post-9/11 symbolic-attacker LAs, we have found four common characteristics:
\begin{itemize}
\item 72.9\% of the symbolic attackers use other weapons than firearms.
\item 70.3\% of their attacks end up with no injuries or fatalities.
\item 67.5\% of symbolic attackers do not have prior contacts to extremist/terrorist groups.
\item 64.5\% of symbolic attackers require a triggering event for the attack intent. 
\end{itemize}

\noindent We have found two strong two-way associations among symbolic attackers:
\begin{itemize}
\item Self-radicalization implies no prior contacts to extremist/terrorist groups, and vice versa. 
\item The usage of other weapons than firearms implies no casualties, and vice versa.
\end{itemize}

\noindent Strong one-way associations found by the \textit{A-priori} algorithm are (Figure \ref{fig:figure10}):

\begin{itemize}
\item No casualties are implied by the existence of a triggering event or being self-radicalized. 
\item A triggering event implies not having prior contacts to extremist/terrorist groups.
\item Self-radicalization implies a trigger event.
\end{itemize}

\subsubsection{Association Rules for Daredevils}

Among 27 post-9/11 daredevil LAs, we have found five common characteristics:
\begin{itemize}
\item 88.9\% of the daredevils use firearms.
\item 74.1\% of the daredevils require a triggering event for the attack intent. 
\item 70.4\% of the daredevils target the law enforcement or government officials.
\item 70.4\% of the daredevils do not leak the attack intent. 
\item 62.9\% of their attacks are fatal. 
\end{itemize}

\noindent We have found four strong two-way associations among daredevils:
\begin{itemize}
\item Having no prior contacts to extremist/terrorist groups implies no leakage, and vice versa. 
\item A triggering event implies targeting the law enforcement or government officials, and vice versa.
\item A triggering event implies targeting the usage of firearms officials, and vice versa.
\item The usage of firearms implies targeting the law enforcement or government officials, and vice versa.
\end{itemize}

\noindent Strong one-way associations found by the \textit{A-priori} algorithm are (Figure \ref{fig:figure11}):
\begin{itemize}
\item Fatalities imply the usage of firearms.
\item A trigger event is implied by not-being self-radicalized.
\item Targeting the law enforcement or government officials is implied by either not-being self-radicalized or fatalities after attack. 
\item The usage of firearms implies not being self-radicalized.
\end{itemize}

\subsubsection{Association Rules for Attention Seekers}

Among 27 post-9/11 attention-seeker LAs, we have found three common characteristics:
\begin{itemize}
\item 67.0\% of the attention seekers target civilians.
\item 62.9\% of their attacks end up with no injuries or fatalities.
\item 62.9\% of the attention seekers do not leak the attack intent. 
\end{itemize}

\noindent No two-way strong associations are found for attention seekers but the following strong one-way associations are found \ref{fig:figure12}):
\begin{itemize}
\item Civilian targets are implied by either of the following: no prior contacts to extremist/terrorist groups, the existence of a triggering event, or usage of other weapons than firearms.
\item The existence of a triggering event is implied by either no prior contacts to extremist/terrorist groups or usage of firearms.
\item Not having prior contacts to extremist/terrorist groups implies no leakage.
\item Usage of other weapons than firearms implies no casualties.
\end{itemize}

\begin{sidewaysfigure}
  \centering
  \begin{subfigure}{0.5\hsize}
    \includegraphics[width=\hsize]{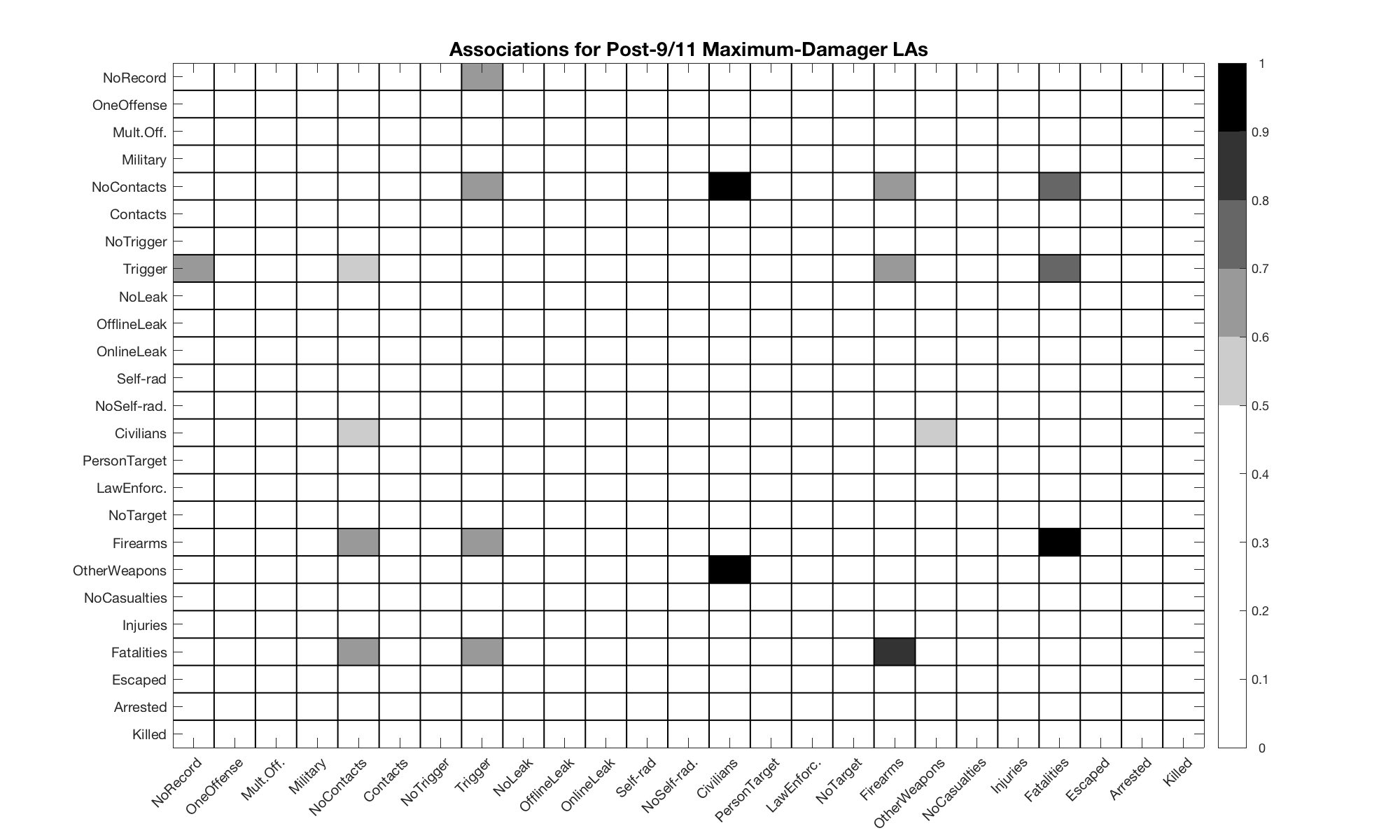}
    \caption{Maximum Damagers}
    \label{fig:figure9} 
  \end{subfigure}%
  \begin{subfigure}{0.5\hsize}
    \includegraphics[width=\hsize]{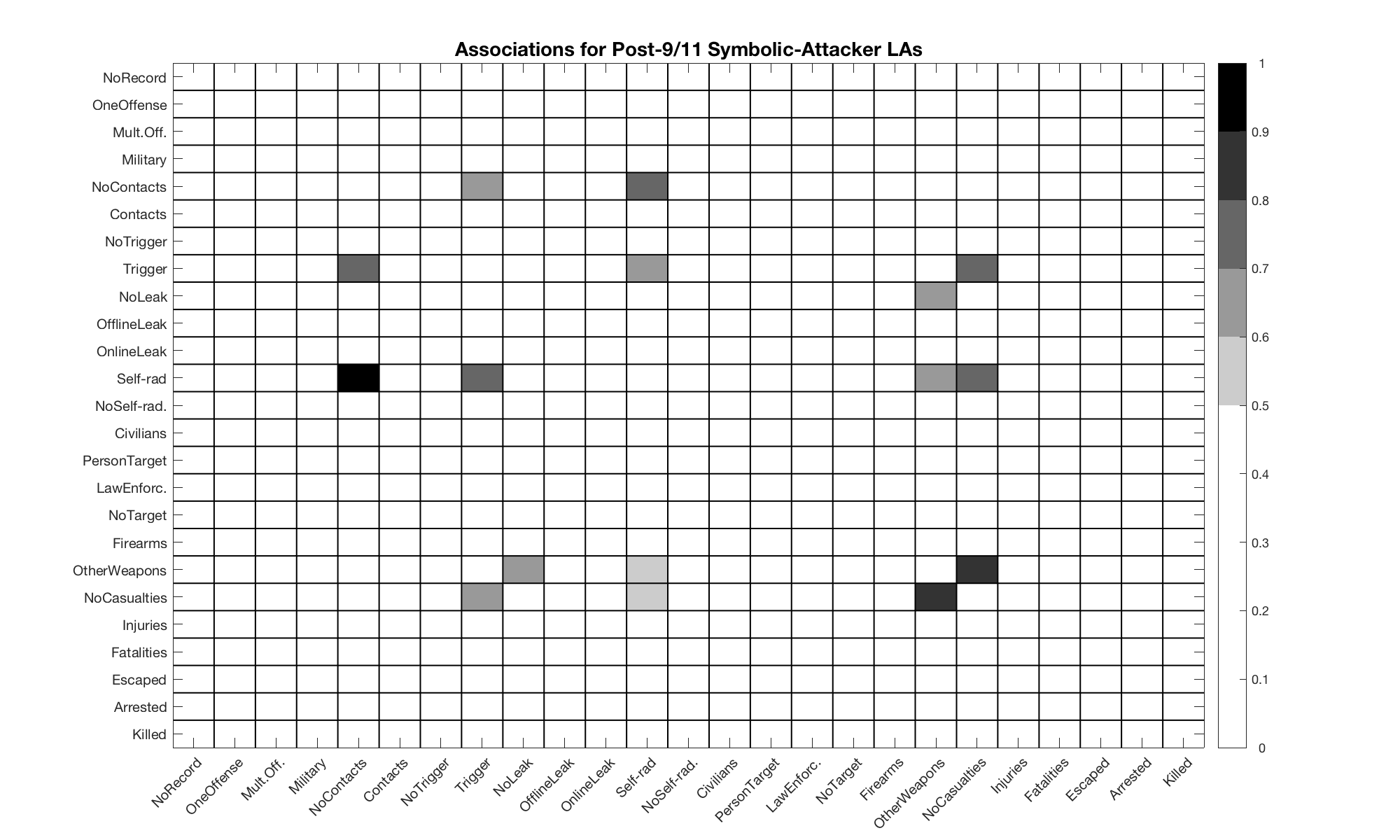}
    \caption{Symbolic Attackers}
    \label{fig:figure10} 
    \end{subfigure}
    
    \begin{subfigure}{0.5\hsize}
    \includegraphics[width=\hsize]{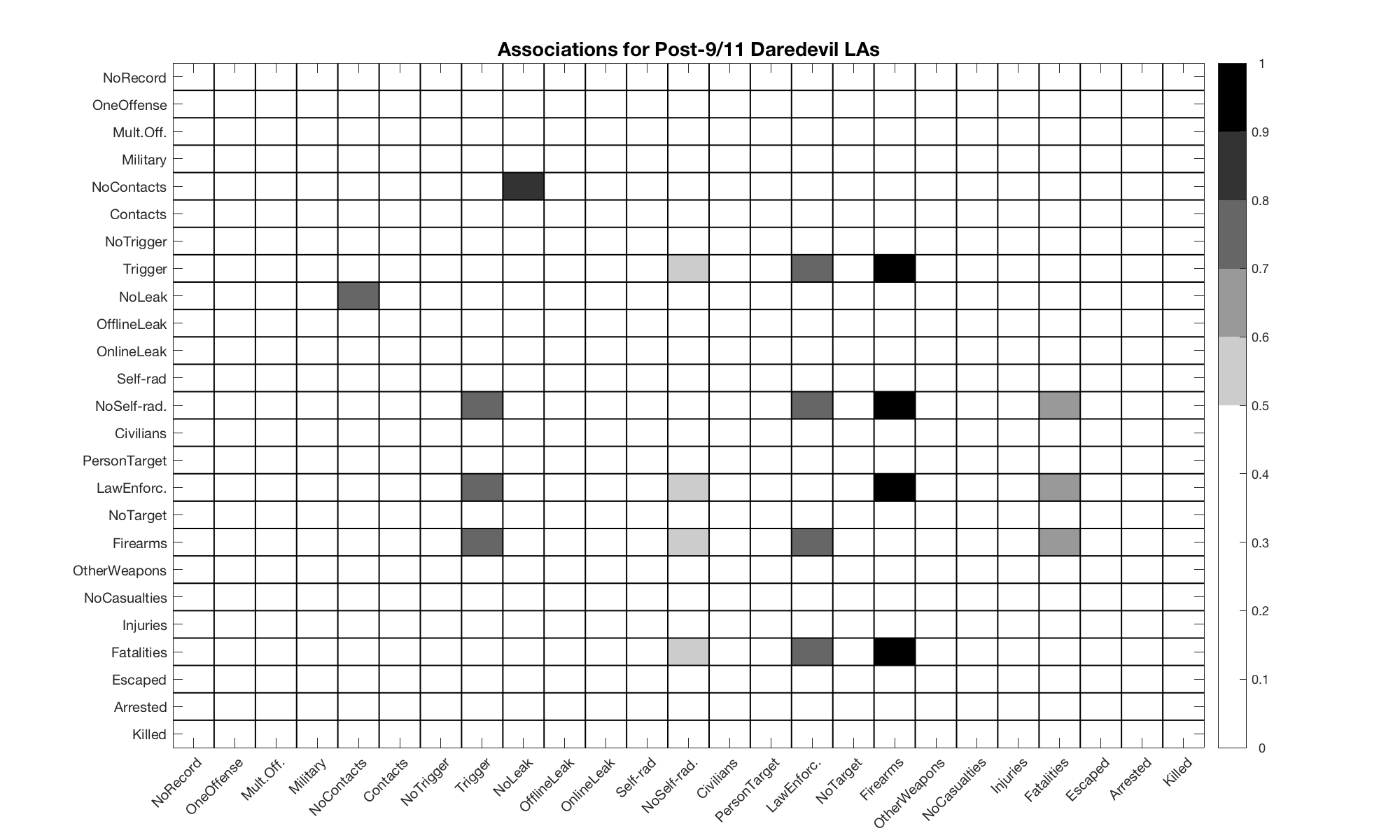}
    \caption{Daredevils}
    \label{fig:figure11} 
    \end{subfigure}%
    \begin{subfigure}{0.5\hsize}
    \includegraphics[width=\hsize]{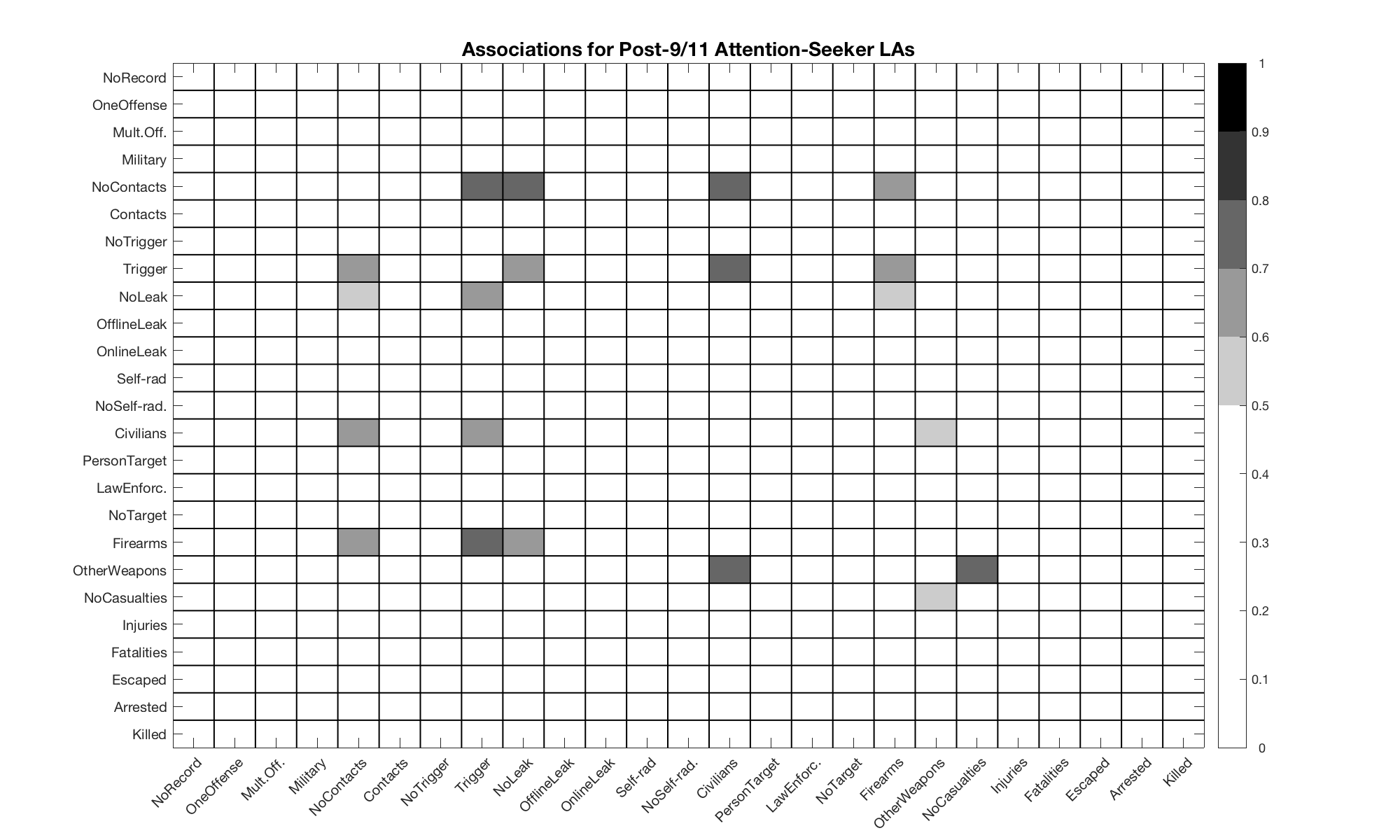}
    \caption{Attention Seekers}
    \label{fig:figure12} 
    \end{subfigure}
    
  \caption{Incident-scene behavior produces more commonalities than ideology-based behavior as can be seen in Table \ref{tab:figure13}. Even though common characteristics have increased in number, most of these commonalities belong to the after-attack behavior. Traceable warning behaviors such as intent leakage or triggering event are not prominent in this classification. \label{fig:incidentb}}
\end{sidewaysfigure}

\subsection{Evaluations regarding behavioral clusters}\label{behavioralc}

This classification method attempts to maximize common characters in behaviors. For this reason, a clustering algorithm (K-Means algorithm) is applied to 152 post-9/11 LA data points. K-Means algorithm requires a predetermined number of clusters. To decide the optimal number of clusters, C-Index is used to validate the clustering results. A smaller C-Index indicates a better clustering~\cite{hubert1976general}. In our trials, seven clusters produce the smallest C-Index value. However, three of these clusters only contain outliers and do not provide sufficient number of data for associations. Hence, we will proceed with four behavioral clusters. 

Table \ref{tab:figure18} shows the distinctive characteristics of each cluster. A triggering event appears as a characteristic for three clusters, and leakage is a common characteristic for one of the clusters. Noticeable similarities are found between the incident-scene-based and behavior-based classification (see Table \ref{tab:table4}). More than half of the first and the third cluster is composed of maximum damagers. 55\% of the LAs in the first cluster are maximum damagers. The LAs in this cluster target civilians, use firearms, and have high fatality rates. 61\% of the fourth cluster is also composed of maximum damagers. The stories and backgrounds of these maximum damagers show that the first cluster contains LAs with a ``me vs. them'' mentality and the fourth cluster contains LAs with a ``us vs. them'' mentality. An LA with a ``me vs. them'' mentality sees every civilian as a target; whereas, an LA with an ``us vs. them'' mentality wants to spare people with their own ideologies and only target civilians from other ideologies. Furthermore, having a ``me vs. them'' mentality increases the fatality rate of attacks. An interesting result is that most ``us vs. them'' maximum damagers have prior contacts to extremist groups and are not self-radicalized. 67\% of the second cluster is composed of symbolic attackers and attention seekers, who appear as self-radicalized. Slightly less than half of the third cluster is composed of daredevils. Since their actions are not well-planned, their attacks are likely to fail and their arrest rate is high.  The second cluster shows some similarity to symbolic attackers who are the majority in the cluster compared to other incident-scene types (48\%). They are self-radicalized LAs with no prior extremist connections and they use other weapons than firearms. Daredevils form 43\% of the third cluster who target the law enforcement and government officials with firearms. They do not leak intent because the duration between trigger event and attack is mostly very short. The last cluster has no triggering event and do not leak intent. Fortunately, their attacks are mostly not fatal. Associations for each cluster is presented in the following subsection.

Table~\ref{tab:aratable} shows the ideological tendencies of each cluster. While first two clusters have a majority of right-wing terrorists, the third cluster (mostly daredevils) is very diverse in terms of ideology. Finally, the last cluster (``us vs. them'' maximum damagers) is almost equally dominated by jihadists and right-wing terrorists.

\begin{table}[ht]
\caption{Cluster Comparisons.}
\centering
\begin{tabular}{cc}
\includegraphics[scale=0.18]{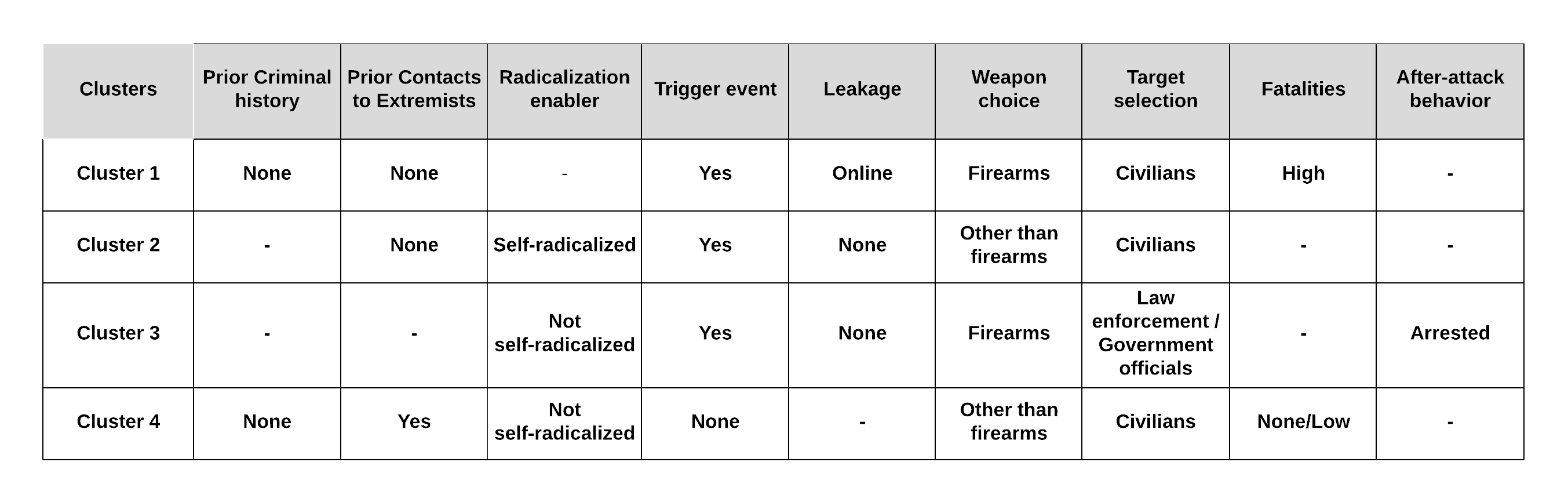}
\end{tabular}
\label{tab:figure18}
\end{table}
 
\begin{table}[]
\begin{tabular}{|c|c|c|c|c|c|}
\hline
          & \begin{tabular}[c]{@{}c@{}}Maximum \\ Damagers\end{tabular} & \begin{tabular}[c]{@{}c@{}}Symbolic \\ Attackers\end{tabular} & Daredevils                  & \begin{tabular}[c]{@{}c@{}}Attention \\ Seekers\end{tabular} & Stallers \\ \hline
Cluster 1 & \textbf{55\%}   & 16\%  & 13\% & 13\%  & 3\%       \\ \hline
Cluster 2 & 23\%   & \textbf{48\%}  & 6\% &\textbf{19\%}   & 3\% \\ \hline
Cluster 3 & 9\%   & 17\% & \textbf{43\%} & 30\%    & 0\%       \\ \hline
Cluster 4 & \textbf{61\%}   & 22\%   & 0\%   & 17\%  & 0\%       \\ \hline
\end{tabular}
\caption {Comparison of two clustering domains}
\label {tab:table4}
\end{table}

\begin{table}[]
\begin{tabular}{|l|c|c|c|c|}
\hline
             & \multicolumn{1}{l|}{Cluster 1}                       & \multicolumn{1}{l|}{Cluster 2} & \multicolumn{1}{l|}{Cluster 3} & \multicolumn{1}{l|}{Cluster 4} \\ \hline
Jihadists    & 16.1\%                                                & 19.4\%                          & 37.5\%  & \textbf{47.8\%}  \\ \hline
Right-wing   &\textbf{51.6\%} & \textbf{48.4\%}  & 29.2\%  &  \textbf{43.5}\%  \\ \hline
Single-issue & 16.1\%                                                & 19.4\%                          & 29.2\%                          & 4.3\%                           \\ \hline
\end{tabular}
\caption {Relationship between ideology-based and behavioral-based classification}
\label {tab:aratable}
\end{table}

\subsubsection{Association Rules for Cluster 1}

Cluster 1 contains 31 LAs. Among these LAs, the common characteristics we have found are:
\begin{itemize}
\item 100.0\% of them target civilians.
\item 96.7\% of them use firearms.
\item 90.3\% of them require a triggering event for the attack intent.
\item 87.1\% of their attacks are fatal.
\item 77.4\% of them have no prior contacts to extremist/terrorist groups.
\item 65.4\% of them commit suicide or are killed at the crime scene.
\end{itemize}

\noindent We have identified six strong two-way associations:
\begin{itemize}
\item Not having prior contacts to extremist/terrorist groups implies civilian targets, and vice versa.
\item Online leakage implies committing suicide or being killed at the crime scene, and vice versa.
\item Not having prior contacts to extremist/terrorist groups implies fatalities, and vice versa.
\item Civilian targets imply a triggering event, and vice versa. 
\item The usage of firearms implies civilian targets, and vice versa.
\item The usage of firearms implies fatalities, and vice versa.
\end{itemize}

\noindent Additionally, we have found the following strong one-way associations:
\begin{itemize}
\item Civilian targets are implied by either of the following: no prior criminal history, the existence of a triggering event, online leakage, or committing suicide or being killed at the crime scene.
\item Not having prior criminal history implies no prior contacts to extremist/terrorist groups and fatalities.
\item Committing suicide or being killed at the crime scene implies fatalities.
\end{itemize}

\subsubsection{Association Rules for Cluster 2}

Cluster 2 contains 27 LAs. Among these LAs, the common characteristics we have observed are:
\begin{itemize}
\item 93.6\% of them have no prior contacts to extremist/terrorist groups.
\item 93.6\% of them use other weapons than firearms.
\item 87.1\% of them are self-radicalized.
\item 77.2\% of them do not leak the attack intent. 
\item 67.8\% of them require a triggering event for the attack intent.
\item 64.5\% of them target civilians. 
\item 64.5\% of their attacks have no casualties. 
\end{itemize}

\noindent We have found three strong two-way associations:
\begin{itemize}
\item The existence of a triggering event implies civilian targets, and vice versa.
\item No leakage implies not having any prior contacts to extremist/terrorist groups, and vice versa.
\item Self-radicalization implies not having any prior contacts to extremist/terrorist groups, and vice versa.
\end{itemize}
\noindent We have also found the following strong one-way associations:
\begin{itemize}
\item No casualties imply the existence of a triggering event, not having any prior contacts to extremist/terrorist groups, and self-radicalization.
\item Not having prior criminal history implies no leakage and self-radicalization.
\item Civilian targets imply no leakage and usage of other weapons than firearms.
\end{itemize}

\subsubsection{Association Rules for Cluster 3}

Cluster 3 contains 24 LAs. The common characteristics of these LAs are given below:
\begin{itemize}
\item 91.7\% of them are not self-radicalized. 
\item 87.5\% of them require a triggering event for the attack intent.
\item 83.3\% of them have prior contacts to extremist/terrorist groups.
\item 79.7\% of them target the law enforcement or government officials.
\item 85.0\% of them use firearms.
\item 70.8\% of them do not leak the attack intent.
\item 87.0\% of their attacks have no casualties.
\item 87.0\% of them are arrested at the crime scene.
\end{itemize}

\noindent We have found five strong two-way associations such that:
\begin{itemize}
    \item Not being self-radicalized implies targeting the law enforcement/government officials, and vice versa.
     \item Not being self-radicalized implies having prior contacts to extremist/terrorist groups, and vice versa.
     \item Not being self-radicalized implies the usage of firearms, and vice versa.
     \item Having prior contacts to extremist/terrorist groups implies targeting the law enforcement/government officials, and vice versa.
     \item The usage of firearms implies targeting the law enforcement/government officials, and vice versa.
\end{itemize}

\noindent We have also found the following strong one-way associations
\begin{itemize}
\item Being arrested after the attack implies no leakage, the usage of firearms, and targeting the law enforcement/government officials.
\end{itemize}

\subsubsection{Association Rules for Cluster 4}

Cluster 4 contains 23 LAs. The common characteristics of these LAs are given below:
\begin{itemize}
\item 100.0\% of them use other weapons than firearms.
\item 87.0\% of them target civilians.
\item 87.0\% of their attacks have no casualties.
\item 87.0\% of them are arrested at the crime scene.
\item 73.9\% of them have prior contacts to extremist/terrorist groups.
\item 60.8\% of them are not self-radicalized. 
\end{itemize}

\noindent We have found seven strong two-way associations: 
\begin{itemize}
\item  Not being self-radicalized implies having prior contacts to extremist/terrorist groups, and vice versa.
\item Civilian targets imply having prior contacts to extremist/terrorist groups, and vice versa.
\item The usage of other weapons than firearms implies having prior contacts to extremist/terrorist groups, and vice versa.
\item The usage of other weapons than firearms implies civilian targets, and vice versa.
\item The usage of other weapons than firearms implies no casualties, and vice versa.
\item The usage of weapons other than firearms implies being arrested at the incident scene, and vice versa.
\item Being arrested at the incident scene implies civilian targets, and vice versa. 
\end{itemize}

\noindent We also found the following strong one-way associations:
\begin{itemize}
\item Having no prior criminal history implies having prior contacts to extremist/terrorist groups, civilian targets, and the usage of weapons other than firearms.
\item Having no trigger event implies contacts to extremist/terrorist groups, the usage of weapons other than firearms, and being arrested at the incident scene.
\item Not being self-radicalized implies the usage of weapons other than firearms.
\end{itemize}

\begin{sidewaysfigure}
  \centering
  \begin{subfigure}{0.5\hsize}
    \includegraphics[width=\hsize]{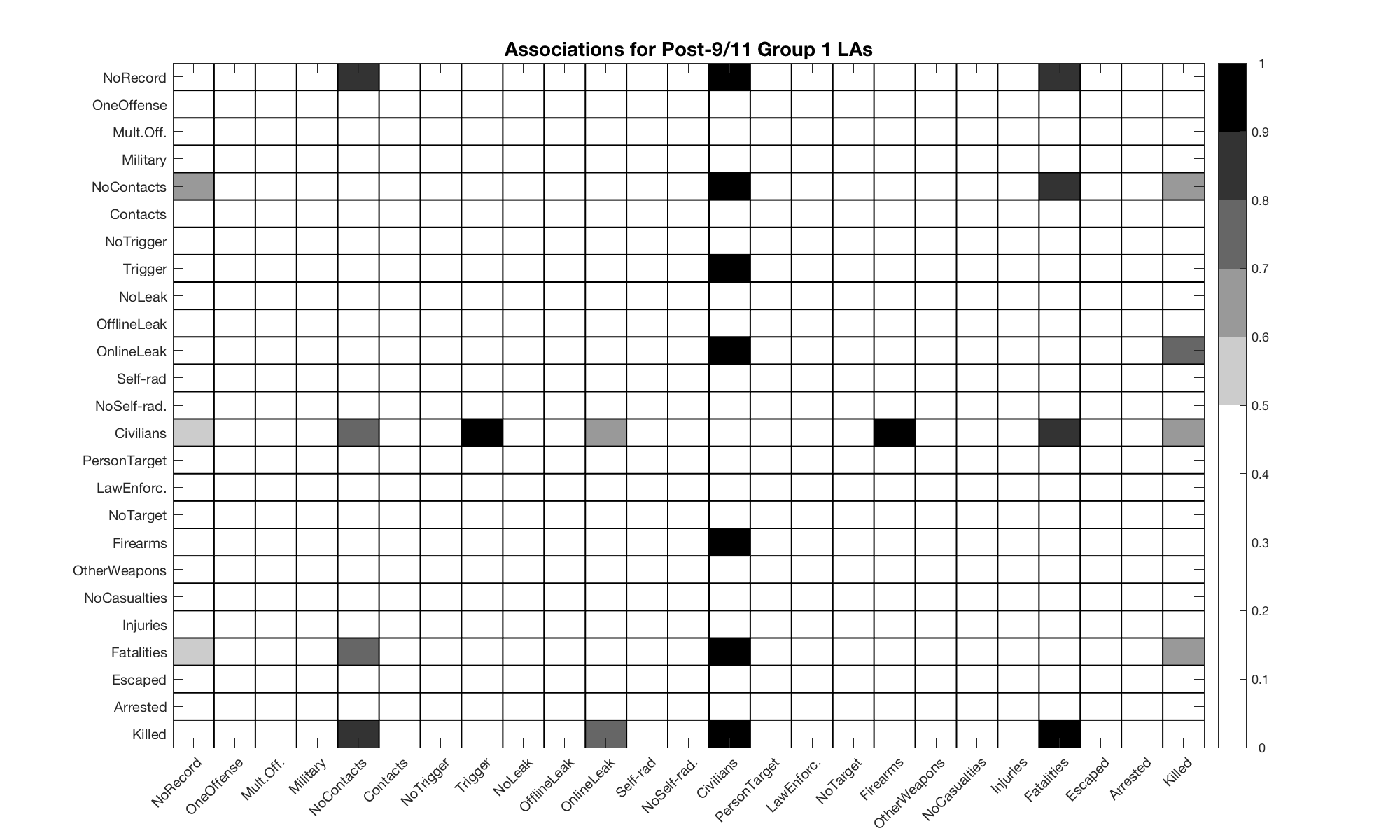}
    \caption{Cluster 1}
    \label{fig:figure14} 
  \end{subfigure}%
  \begin{subfigure}{0.5\hsize}
    \includegraphics[width=\hsize]{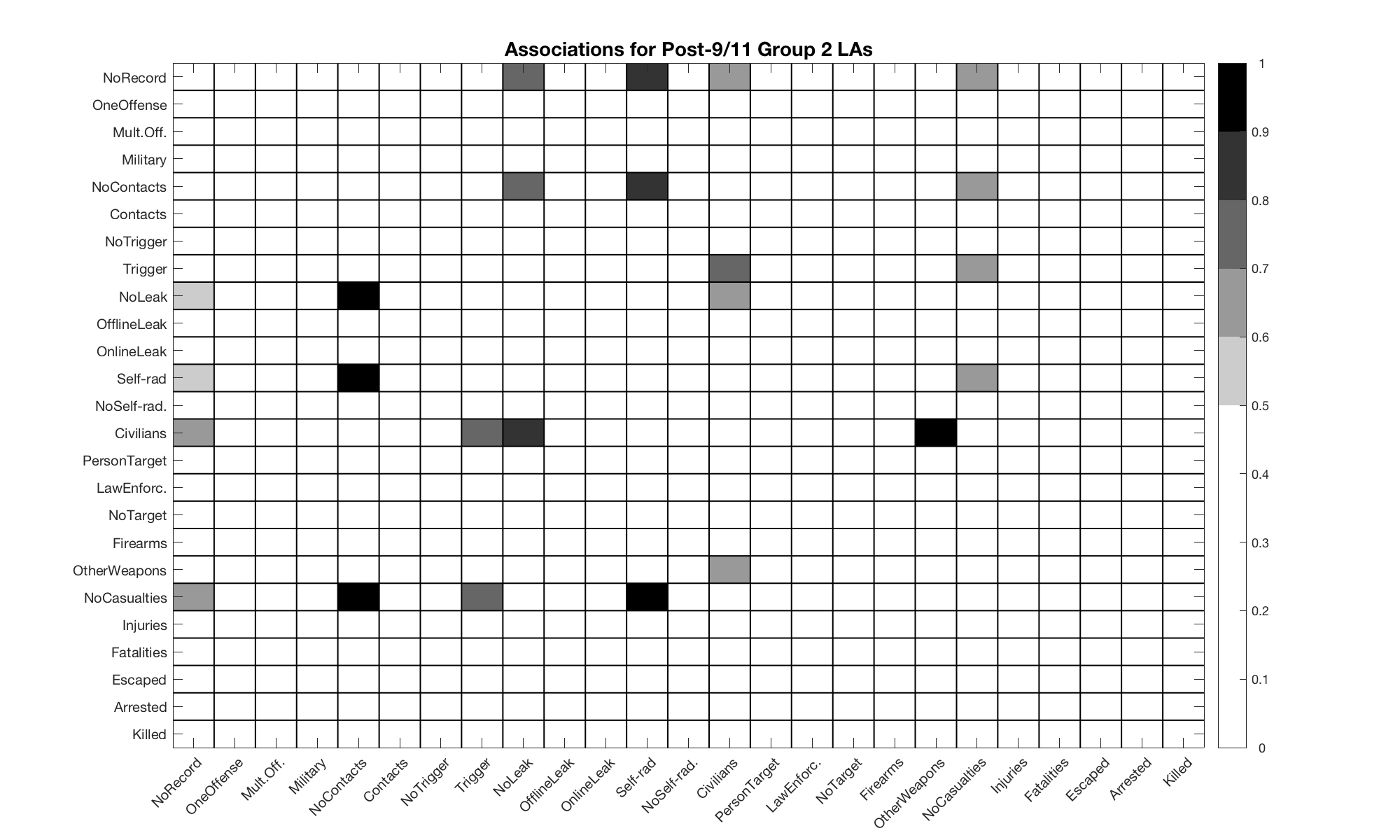}
    \caption{Cluster 2}
    \label{fig:figure15} 
    \end{subfigure}
    
    \begin{subfigure}{0.5\hsize}
    \includegraphics[width=\hsize]{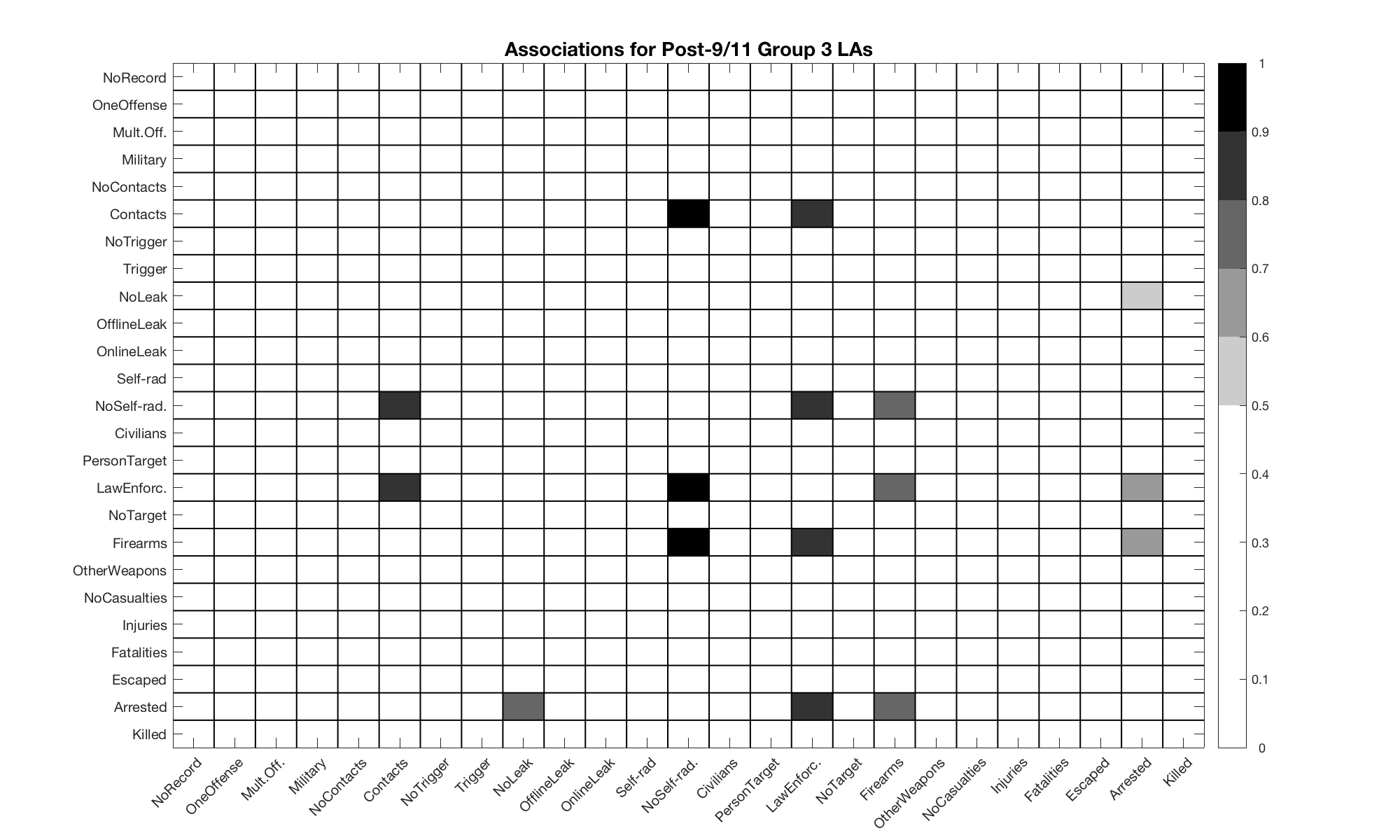}
    \caption{Cluster3}
    \label{fig:figure16} 
    \end{subfigure}%
    \begin{subfigure}{0.5\hsize}
    \includegraphics[width=\hsize]{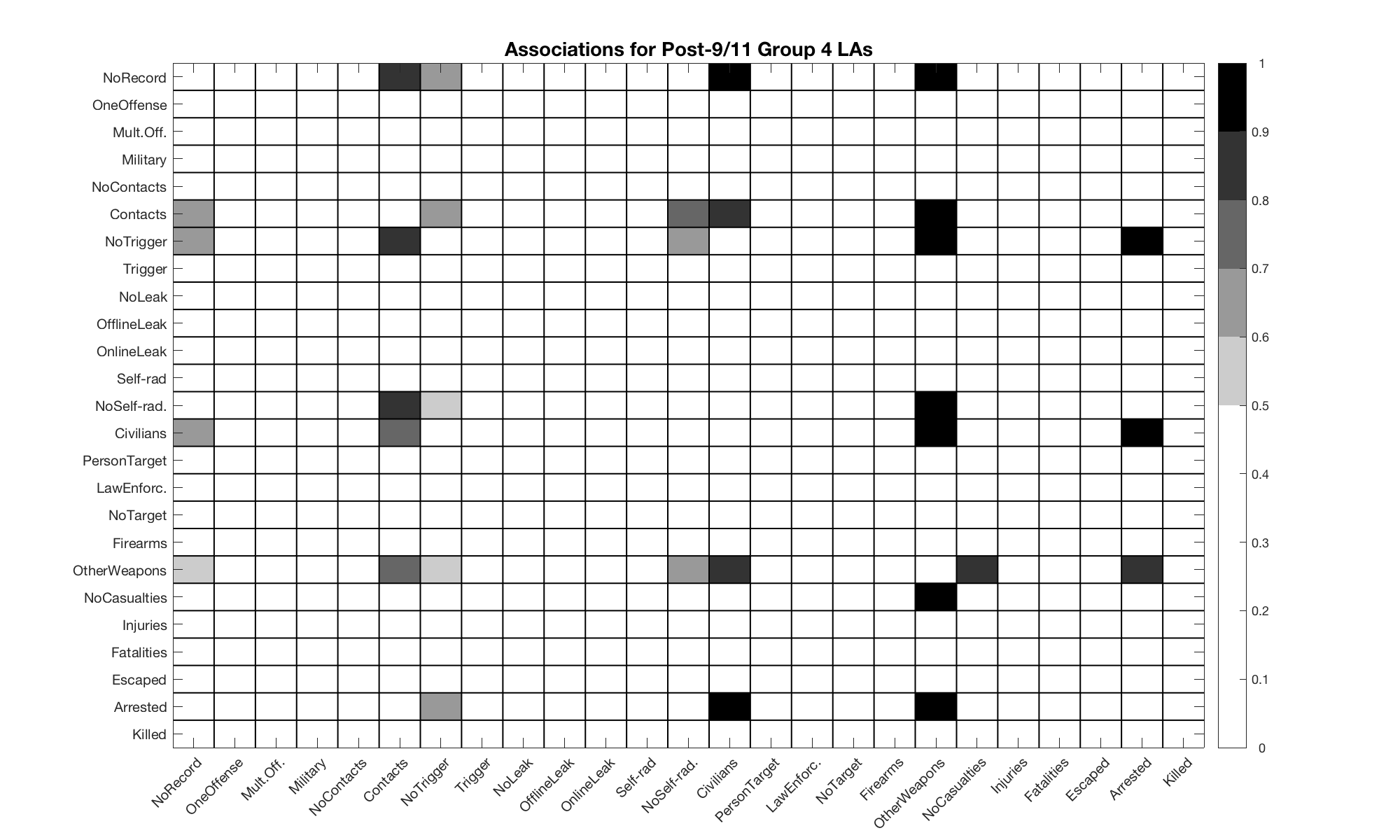}
    \caption{Cluster4}
    \label{fig:figure17} 
    \end{subfigure}
    
  \caption{Behavior-based classifications produces considerably more commonalities and associations than previous classifications. \label{fig:behaviorb}}
\end{sidewaysfigure}

\section{Association Chains} \label{chains}

In this section, we extract chain rules. A chain rule is a chain of associations that satisfy a cohesion threshold level of 70\% and an overall minimum confidence threshold level of 70\% given an initial node. Each node represents a behavior or an attack characteristic and is filled with a color code indicating chronological occurrence. The color codes for the timeline is presented in Figure \ref{fig:figure19}. Finally, two-way arrows indicate that if the nodes on each edge of a two-way arrow are interchanged, the chain still satisfy the minimum confidence level for that rule.

\begin{figure}[H]
	\includegraphics[width=\textwidth]{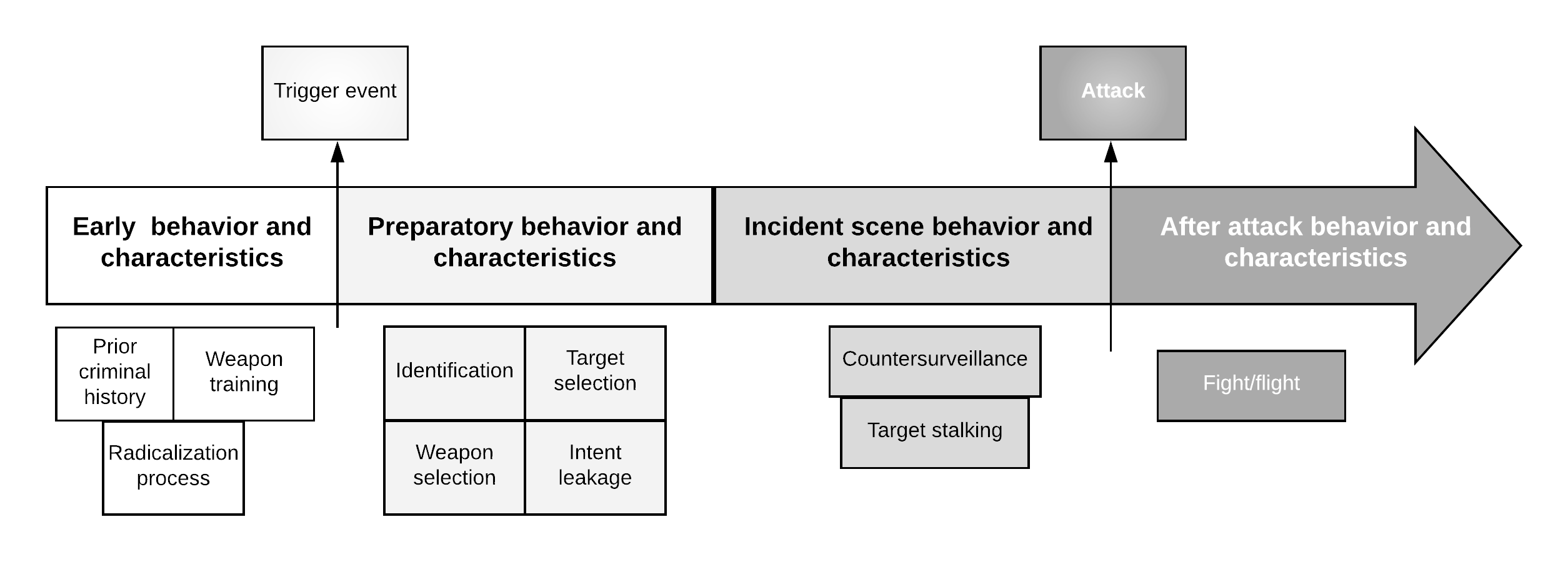}
	\caption{Timeline color codes for chain rules. The lighter shade of cells indicate an event in the further past. The darker shades indicate closeness to an attack.}
	\label{fig:figure19}
\end{figure}

Overall evaluation of LAs and post-9/11 LAs do not produce any implication chains. However, Figure \ref{fig:figure20} presents the chains for pre-9/11 LAs. The central theme for pre-9/11 LAs is not being self-radicalized, fatalities, and the usage of firearms. As aforementioned, while pre-9/11 LAs are relatively homogeneous, post-9/11 LAs are very heterogeneous and do not produce any association chains. Similarly, ideological classification has not produced any association chains, as well. 

\begin{figure}[H]
	\includegraphics[scale=0.17]{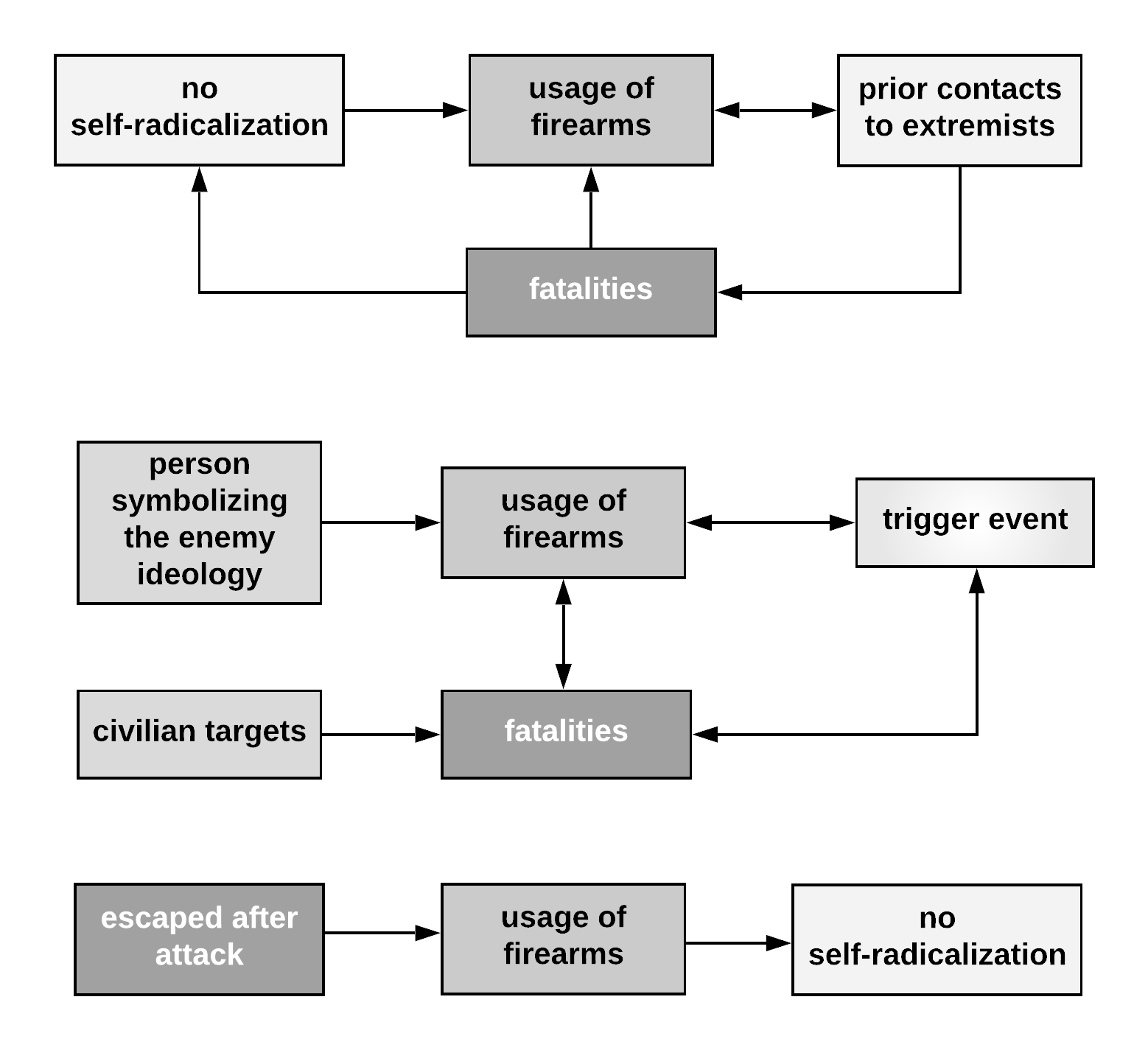}
	\caption{Chain Rules for Pre-9/11 LAs.}
	\label{fig:figure20}
\end{figure}

\noindent Among the incident-scene behavior-based classification, maximum damagers and attention seekers do not yield any chains. However, the chains for symbolic attackers and daredevils are given in Figure \ref{fig:figure21} and Figure \ref{fig:figure22}. 

\begin{figure}[H]
	\includegraphics[scale=0.2]{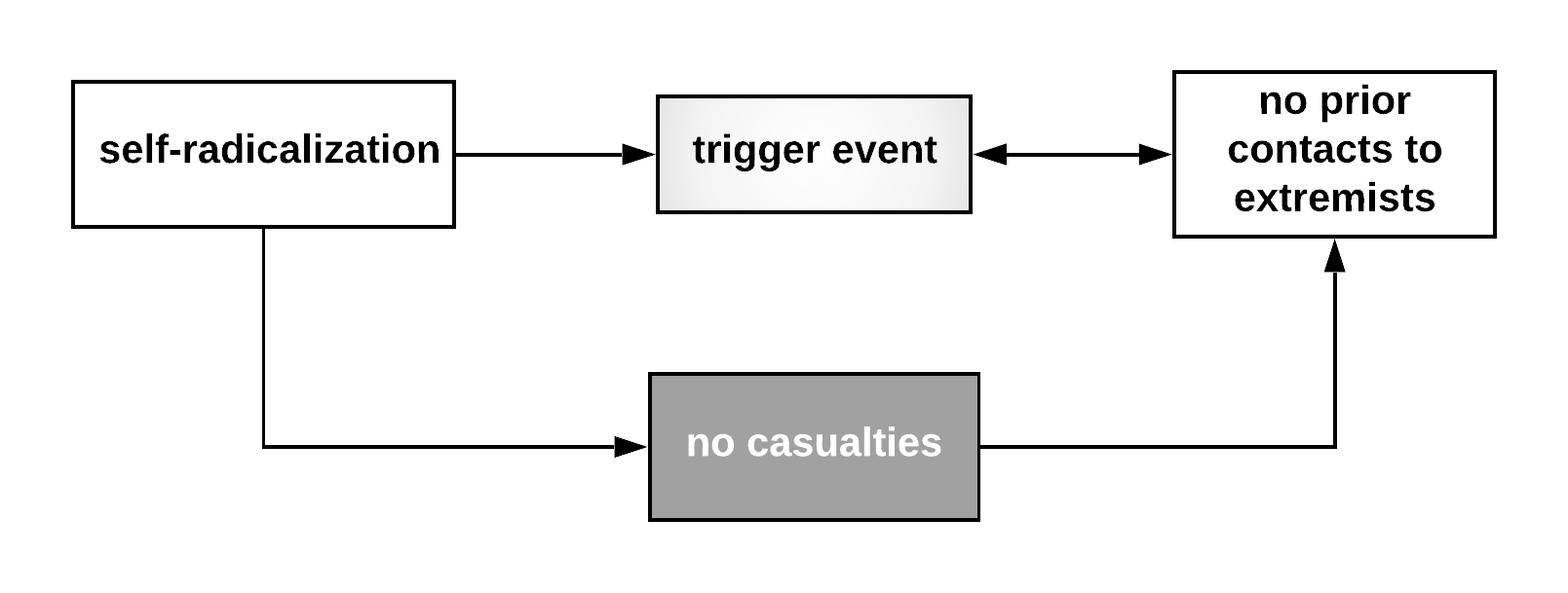}
	\caption{Chain Rules for Symbolic Attackers.}
	\label{fig:figure21}
\end{figure}

\begin{figure}[H]
	\includegraphics[scale=0.2]{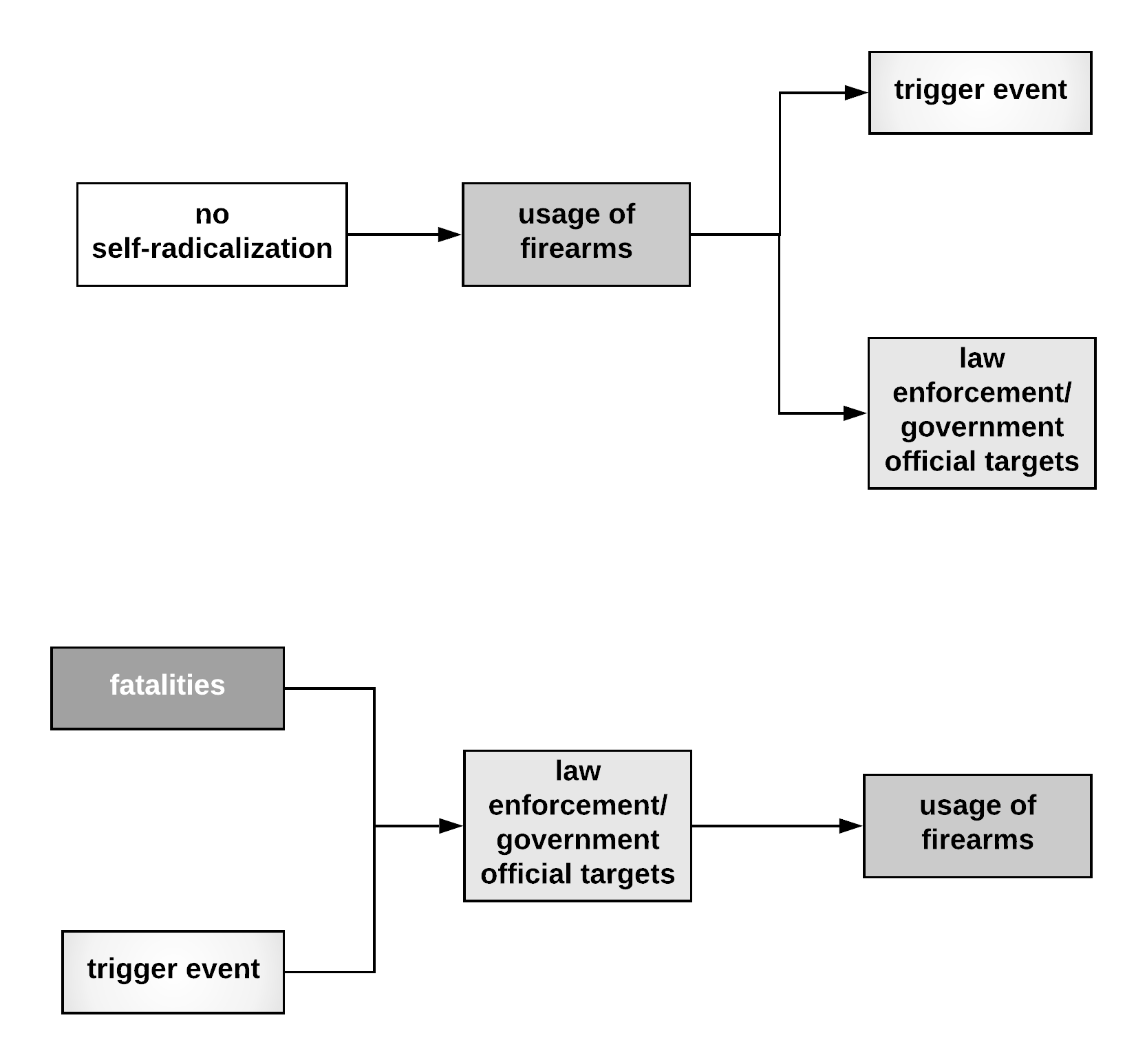}
	\caption{Chain Rules for Daredevils.}
	\label{fig:figure22}
\end{figure}

While symbolic attackers are mostly self-radicalized, their attacks do not necessarily aim to kill. Hence, low casualties appear as the central theme. Daredevils, on the other hand, are mostly not self-radicalized. They do not generally leak intent because their attacks are the most impulsive ones among LAs. They either target the law enforcement face-to-face or exhibit violence at the same moment as the trigger. Their attacks are mostly fatal. 

Behavior-based classification has yielded longer and statistically more robust chains compared to other classification schemes. The association chain for Cluster 1 is presented in Figure \ref{fig:figure23}. This group involves more ordinary people who may not be considered threatening. They have no prior criminal history or contacts to other extremist, radical, or terrorist groups. However, after a trigger, their grievance channels to an attack intent with the use of firearms. Consequently, fatality rate of their attacks is high. Cluster 1 is the only group that leaks intent and their leakage pattern is mostly online. Hence, they leave their ``writeprints'' on the internet.

\begin{figure}[H]
	\includegraphics[scale=0.2]{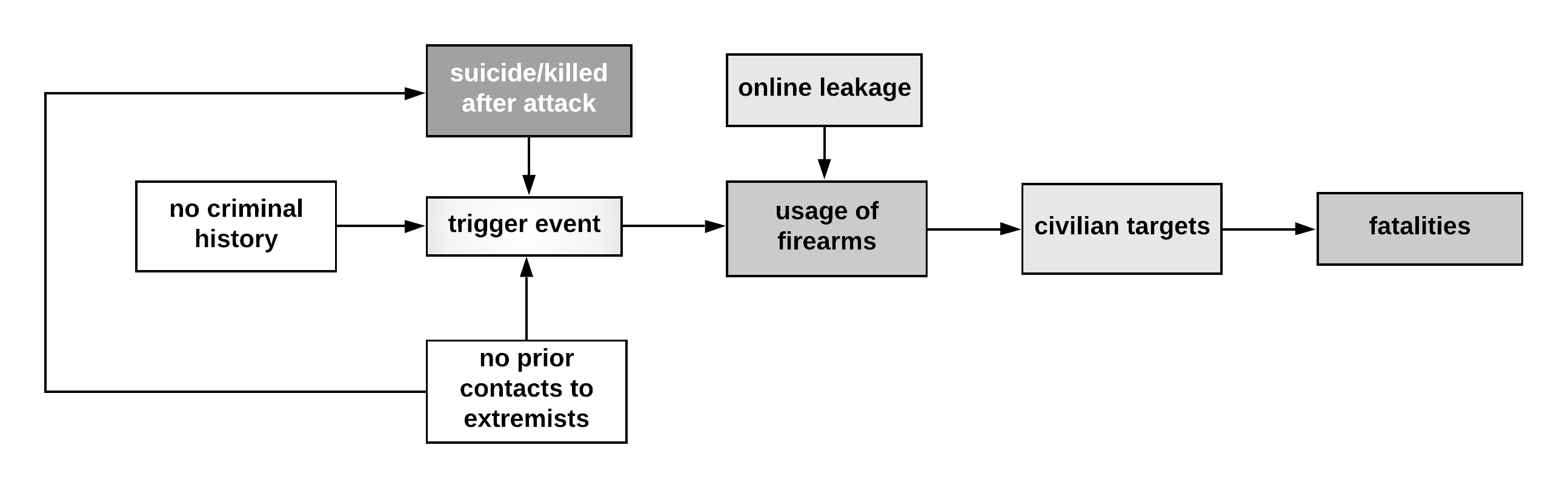}
	\caption{Chain Rules for Cluster 1.}
	\label{fig:figure23}
\end{figure}

Figure \ref{fig:figure24} exhibits the association chain for Cluster 2. The attackers in this cluster do not usually use firearms. They have no prior contacts to other extremist, radical, or terrorist groups, and they are self-radicalized. They experience a triggering event similar to the ones in Cluster 1, however, unlike Cluster 1, they do not leak intent. Moreover, they mostly target civilians.
\begin{figure}[H]
	\includegraphics[scale=0.2]{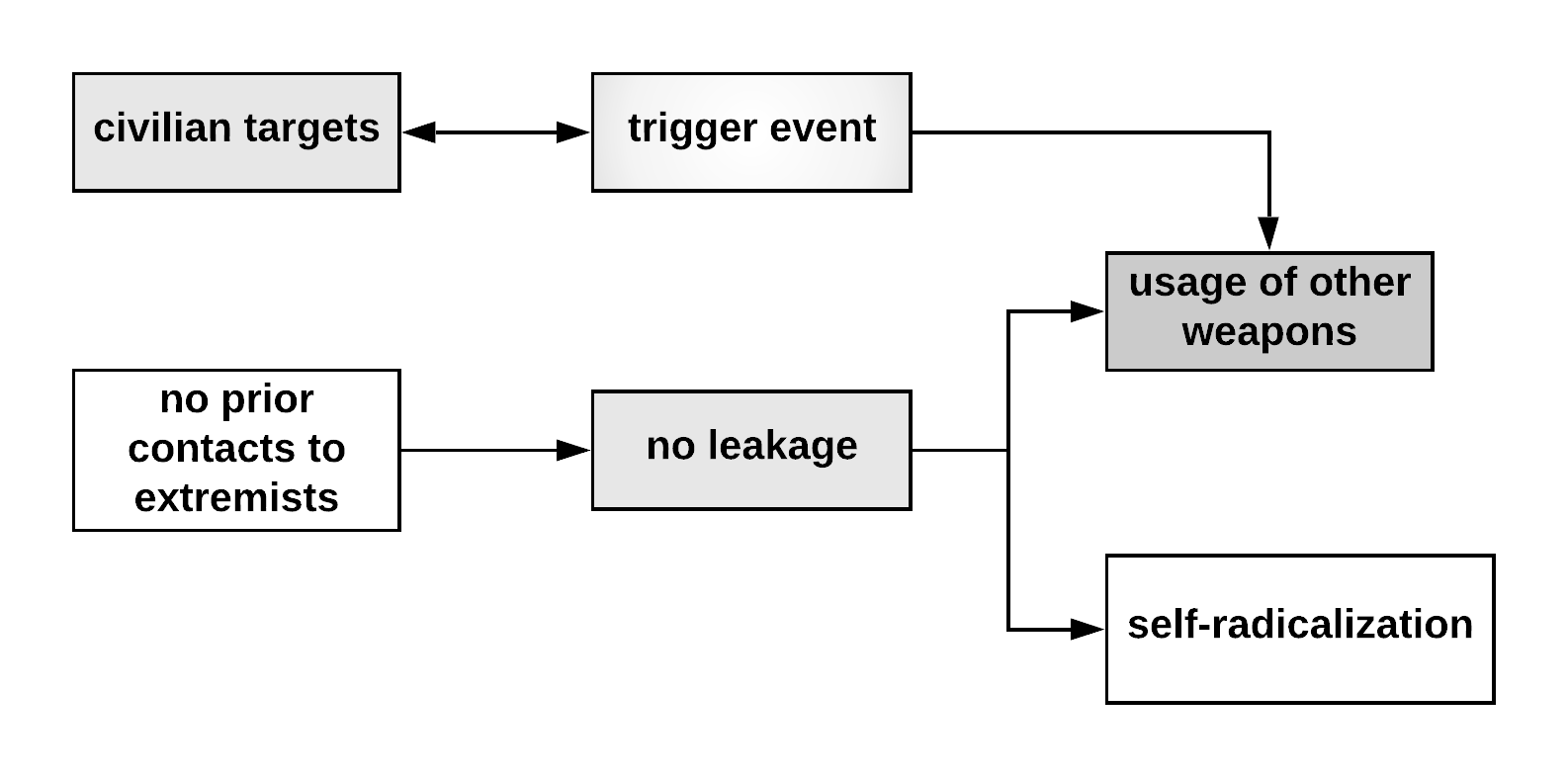}
	\caption{Chain Rules for Cluster 2.}
	\label{fig:figure24}
\end{figure}

Figure \ref{fig:figure25} demonstrates the association chain for Cluster 3. This cluster exhibits similar behavior to daredevils; their targets are the law enforcement or government officials, and they use firearms. They are mostly arrested after the attack. While ``no leakage'' is an important characteristic of this cluster; a trigger event is not prominent, as well.

\begin{figure}[H]
	\includegraphics[scale=0.2]{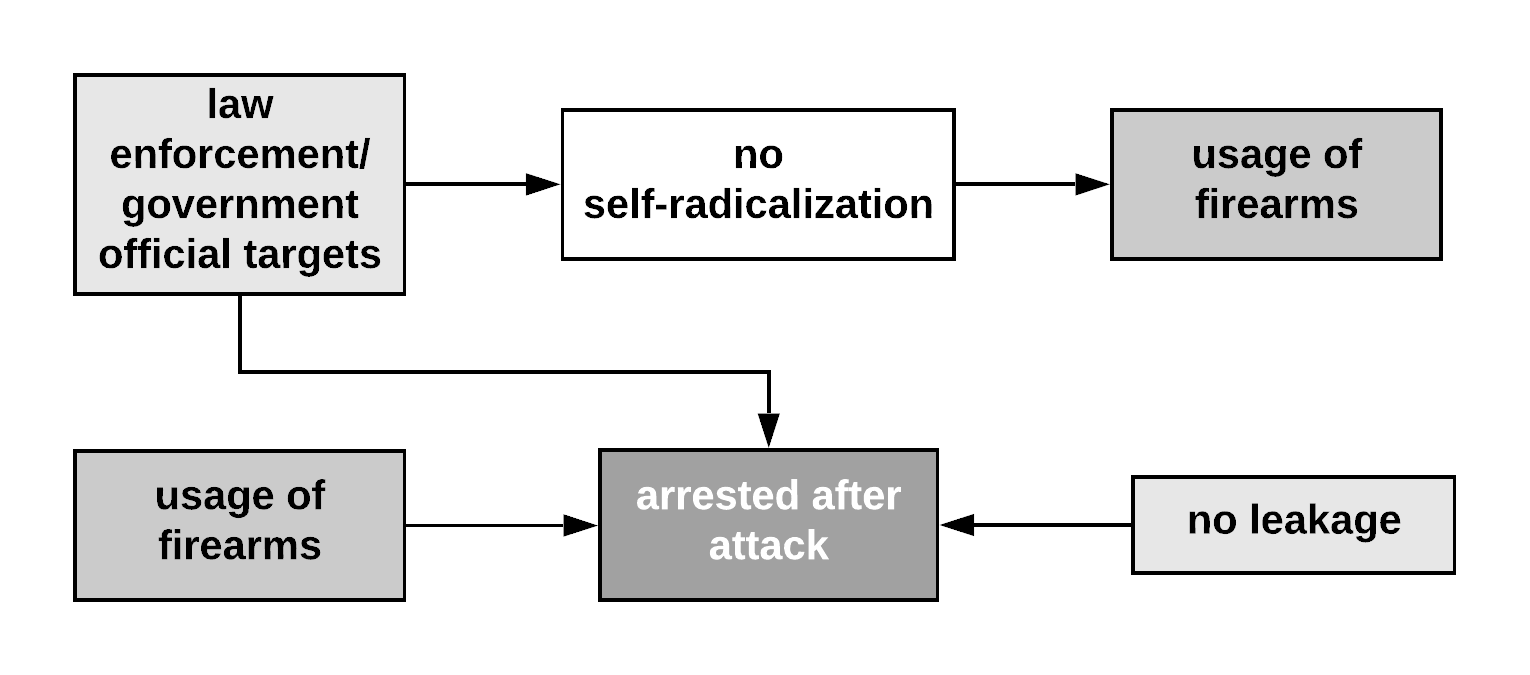}
	\caption{Chain Rules for Cluster 3.}
	\label{fig:figure25}
\end{figure}

Figure \ref{fig:figure26} displays the association chain for Cluster 4. The LAs in this cluster are not subjected to a trigger event and have no prior criminal history. In contrast to Cluster 2, they have prior contacts to other extremist, radical, or terrorist groups and they are not self-radicalized. However, similar to Cluster 2, they also choose weapons other than firearms; hence, rate of fatal attacks is low.  

\begin{figure}[H]
	\includegraphics[scale=0.2]{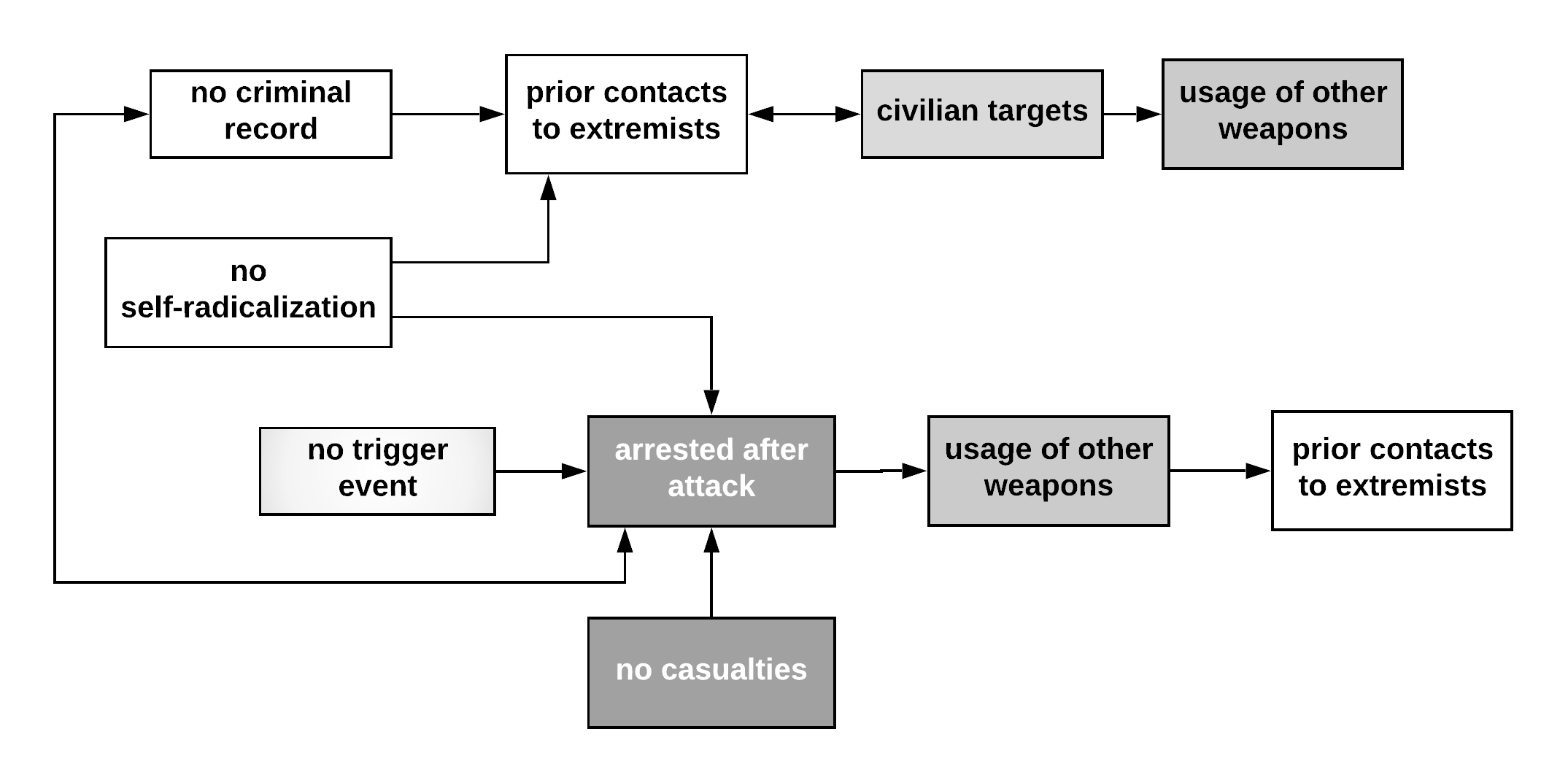}
	\caption{Chain Rules for Cluster 4.}
	\label{fig:figure26}
\end{figure}

\section{Temporal Associations} \label{tempAs}

Among behavioral characteristics, online leakage and trigger events are important and traceable milestones on the pathway to an attack. In this section, we will provide statistical properties of each LA type. The most prominent landmark is the trigger event and its timing.  Some of the timings are available on the databases. However, for most of these timings,  we were able to gather information through local and national newspaper archives. Pre-9/11 LAs are excluded in the analyses because their trigger times are seldom readily available online. 

72 of the 99 triggering events have the exact dates. 24 attacks have the exact week, two of them have the exact month and four of them have the exact year. In the statistical analyses, we have used the day and the week data because month and year information assumptions largely affect the results. The statistical properties of the duration between trigger event and the attack are given in Table \ref{tab:table2}. The temporal deviations between triggers for each group are very high and even though the association rules and chains hold, the temporal deviations between trigger event and attack are very large.

\begin{table}[h]
\tiny
\begin{tabular}{|l|c|c|c|c|c|c|}
\hline
                   & \textbf{Number of Data Points} & \textbf{Prominent Characteristic} & \textbf{Mean} &\textbf{Median}  & \textbf{Standard Deviation} & \textbf{Coefficient of Variation} \\ \hline
Overall            & 93                    & Yes                      & 205.7 &31& 419.9              & 2                        \\ \hline
\multicolumn{6} {|c|}{\textbf{Ideology-Based}}                 \\ \hline
Jihadists          & 24                    & No                       & 150  &46 & 320.4              & 2.1                      \\ \hline
Right-wing         & 39                    & Yes                      & 205.2&18 & 366.8              & 1.8                      \\ \hline
Single-issue       & 20                    & Yes                      & 188.5&60 & 370.5              & 2                        \\ \hline
\multicolumn{6} {|c|}{\textbf{Incident-Scene  Based}}               \\ \hline
Maximum Damagers   & 30                    & No                       & 364.5 &151& 553.8              & 1.5                      \\ \hline
Symbolic Attackers & 26                    & Yes                      & 149.2&27 & 291                & 2                        \\ \hline
Daredevils         & 20                    & Yes                      & 192.3 &31& 445.8              & 2.3                      \\ \hline
Attention Seekers  & 15                    & Yes                      & 26.9  &4& 59.6               & 2.2                      \\ \hline
\multicolumn{6} {|c|}{\textbf{Behavior-Based}}                     \\ \hline
Group 1            & 26                    & Yes                      & 298.7 &144 & 388.8              & 1.3                      \\ \hline
Group 2            & 22                    & Yes                      & 41.5 &7& 86.9               & 2.1                      \\ \hline
Group 3            & 19                    & Yes                      & 173.3 &17& 379.6              & 2.2                      \\ \hline
Group 4            & 8                     & No                       & 209.5 &31& 389.2              & 1.9                      \\ \hline
\end{tabular}
\caption {Temporal Statistical Properties between Trigger Event and Attack}
\label {tab:table2}
\end{table}

Another important landmark is the leakage but only one attacker type has leakage as a prominent characteristic. Cluster 1 has scored high on online leakage. In this group, 20 exact leakage dates are found and the mean between leakage and attack is 60 days with a large standard deviation of 87 days. Despite the large standard deviation, the median of the time span between the leakage and the attack is 7 days, meaning that 50\% of the LAs attack within a week after the leakage. 

Intervals between the landmarks show a vast variation; all groups have high coefficient of variation values and the duration cannot be generalized within groups. Another arising question is that if these durations are related with the rate of fatal attacks. Non-fatal attacks have an average of 115 days with a standard deviation of 234.7 days between the trigger and the attack; whereas fatal attacks have an average of 141 days with a standard deviation of 234.8 days between the trigger and the attack. A t-test comparing these groups yields a p-value of 0.63 which indicates that fatal and non-fatal attacks do not have a significant difference between their trigger-attack durations. Another comparison between fatal and non-fatal attacks have been made in terms of the duration between leakage and attack times. Non-fatal attacks have an average of 44 days with a standard deviation of 53.4 days between the leakage and the attack; whereas fatal attacks have an average of 36 days with a standard deviation of 64.2 days between the trigger and the attack. A t-test comparing these groups yields a p-value of 0.711 which indicates no significant difference between leakage and fatality. 

Despite the large standard deviations of temporal difference between mileposts, medians provide valuable insights. Almost 50\% of the LAs attack in a month after the trigger. Even though daredevils are more impulsive than other types, attention seekers have a much smaller median, that is, 50\% of attention seekers attack in less than 5 days after the trigger.  Likely, 50\% of Group 2 (mostly consists of symbolic attackers and attention seekers) attacks in less than a week after the trigger. One important remark is maximum damagers have longer duration than any other group; hence, it can be argued that high fatality rate requires high attack preparation time. 

\section {Conclusions and Future Work}\label{conclusion}

The face of terrorism has dramatically changed, especially over the last two decades. The number of attacks by individuals have been more frequent than ever, making LA terrorism one of the most accelerating man-made threats, especially in the US. As a result, there is a need for academic work to understand LA behavior and characteristics. While understanding LA behavior have been studied qualitatively and quantitatively, the associations and connections between those behavioral characteristics have not been analyzed in a temporal manner for an attempt to intervene at the right time. 

In this study, distal and proximal characteristics of LAs are analyzed together with attack characteristics and after-attack behaviors. However, the term ``behavior'' indicates different concepts in literature. To provide clarity on this term, we have defined four temporal behavior phases: early behavior, preparatory behavior, incident-scene behavior, and after attack behavior. Using 25 binary characteristics, we first compare pre- and post-9/11 LAs. The results indicate that while pre-9/11 LAs had prior contacts to extremist/radical groups and mostly radicalized by the people in their environment, post-9/11 LAs are more diverse. The most noticeable change we have found is that LA terrorism has trickled down to people that do not have prior connections, and the weapon of choice has diverged greatly.

Additionally, besides ideological classification, we introduce two new classifications of LAs: incident-scene-based and behavior-based classification. Incident-scene-based classification is achieved through the data obtained with the game developed in the GRIST Lab and behavior-based classification is obtained through the clustering of 25 binary characteristics. The incident-scene-based data provides five types of attackers: maximum damagers, symbolic attackers, daredevils, attention seekers, and stallers. Behavior-based classification further divides maximum damagers to ``me vs. them'' and ``us vs. them'' types. Through behavior-based classification, we are able understand the evolution process of an LA attacker by producing association chains. 

Triggering event and leakage are traceable characteristics if they have online ``writeprints''. Hence, our further analysis evaluates if these landmarks differ according to the LA type. However, the durations between triggering event and attack, or the durations between leakage and attack have high standard deviations for each LA type. Moreover, these durations do not have a statistically significant effect on the fatal attack rates. Hence, further analysis is required for connecting behavior to attack timeline.   

\bibliography{article1}

\begin{thebibliography}{10}

\bibitem{phillips2011lone}
P.~J. Phillips, ``Lone wolf terrorism,'' {\em Peace Economics, Peace Science
  and Public Policy}, vol.~17, no.~1, 2011.

\bibitem{spaaij2010enigma}
R.~Spaaij, ``The enigma of lone wolf terrorism: An assessment,'' {\em Studies
  in Conflict \& Terrorism}, vol.~33, no.~9, pp.~854--870, 2010.

\bibitem{hinksonj}
J.~Hinkson, ``Learning from las vegass,'' {\em Arena Magazine}, vol.~150,
  pp.~5--6, 2017.

\bibitem{quek2019paso}
N.~Quek, ``{Growing Threat of White Supremacists},'' tech. rep., RSIS
  Commentaries. Singapore: Nanyang Technological University, 2019.

\bibitem{gill2014bombing}
P.~Gill, J.~Horgan, and P.~Deckert, ``Bombing alone: Tracing the motivations
  and antecedent behaviors of lone-actor terrorists,'' {\em Journal of Forensic
  Sciences}, vol.~59, no.~2, pp.~425--435, 2014.

\bibitem{brynielsson2013harvesting}
J.~Brynielsson, A.~Horndahl, F.~Johansson, L.~Kaati, C.~M{\aa}rtenson, and
  P.~Svenson, ``Harvesting and analysis of weak signals for detecting lone wolf
  terrorists,'' {\em Security Informatics}, vol.~2, no.~1, p.~11, 2013.

\bibitem{pantucci2011topology}
R.~Pantucci, ``{A Topology of Lone Wolves: Preliminary Analysis of Lone
  Islamist Terrorists},'' tech. rep., The International Centre for the Study of
  Radicalisation and Political Violence, 03 2011.

\bibitem{bakker2011preventing}
E.~Bakker and B.~De~Graaf, ``Preventing lone wolf terrorism: Some {C}{T}
  approaches addressed,'' {\em Perspectives on Terrorism}, vol.~5, no.~5--6,
  2011.

\bibitem{meloy2011role}
J.~Reid~Meloy, J.~Hoffmann, A.~Guldimann, and D.~James, ``The role of warning
  behaviors in threat assessment: An exploration and suggested typology,'' {\em
  Behavioral Sciences \& the Law}, vol.~30, no.~3, pp.~256--279, 2011.

\bibitem{gartenstein2015radicalization}
D.~Gartenstein-Ross, ``Radicalization: Social media and the rise of
  terrorism,'' tech. rep., Hearing before the US House of Representatives
  Committee on Oversight and Government Reform, Subcommittee on National
  Security 28, 2015.

\bibitem{spaaij2011understanding}
R.~Spaaij, {\em Understanding Lone Wolf Terrorism: Global Patterns, Motivations
  and Prevention}.
\newblock Springer Briefs in Criminology, Springer Netherlands, 2011.

\bibitem{maccauley2008mechanisms}
C.~McCauley and S.~Moskalenko, ``Mechanisms of political radicalization:
  Pathways toward terrorism,'' {\em Terrorism and Political Violence}, vol.~20,
  no.~3, pp.~415--433, 2008.

\bibitem{meloy2017trap}
J.~R. Meloy, {\em TRAP-18: Terrorist Radicalization Assessment Protocol User
  Manual Version 1.0}.
\newblock Washington, DC: Global Institute of Forensic Research., 2017.

\bibitem{bates2012dancing}
R.~A. Bates, ``Dancing with wolves: Today's lone wolf terrorists,'' {\em The
  Journal of Public and Professional Sociology}, vol.~4, no.~1, pp.~1--14,
  2012.

\bibitem{meloy2016lone}
R.~J. Meloy and P.~Gill, ``The lone-actor terrorist and the trap-18,'' {\em
  Journal of Threat Assessment and Management}, vol.~3, no.~1, pp.~37--52,
  2016.

\bibitem{sawyer2017what}
J.~P. Sawyer and J.~Hienz, {\em What Makes Them Do It? Individual-Level
  Indicators of Extremist Outcomes}, ch.~3, pp.~47--61.
\newblock John Wiley \& Sons, Ltd, 2017.

\bibitem{perliger2011social}
A.~Perliger and A.~Pedahzur, ``Social network analysis in the study of
  terrorism and political violence,'' {\em PS: Political Science and Politics},
  vol.~44, pp.~45--50, 2011.

\bibitem{li2018terrorist}
Z.~Li, D.~Sun, B.~Li, Z.~Li, and A.~Li, ``Terrorist group behavior prediction
  by wavelet transform-based pattern recognition,'' {\em Discrete Dynamics in
  Nature and Society}, pp.~1--16, 2018.

\bibitem{gordon2017potential}
T.~J. Gordon, Y.~Sharan, and E.~Florescu, ``Potential measures for the
  pre-detection of terrorism,'' {\em Technological Forecasting and Social
  Change}, vol.~123, pp.~1 -- 16, 2017.

\bibitem{schuurman2018lone}
B.~Schuurman, E.~Bakker, and N.~B. P.~Gill, ``Lone actor terrorist attack
  planning and preparation: A data-driven analysis,'' {\em Psychiatry and
  Behavioral Science}, vol.~63, no.~4, 2018.

\bibitem{ellis2017lone}
C.~Ellis, R.~Pantucci, J.~van Zuijdewijn, E.~Bakker, B.~Gomis, S.~Palombi, and
  M.~Smith, ``{Lone-Actor Terrorism},'' tech. rep., Royal United Service
  Institute for Defense and Security Studies, 2017.

\bibitem{post1985individual}
J.~M. Post, ``Individual and group dynamics of terrorist behavior,'' in {\em
  Psychiatry: The State of the Art Volume 6 Drug Dependence and Alcoholism,
  Forensic Psychiatry, Military Psychiatry} (P.~Pichot, P.~Berner, R.~Wolf, and
  K.~Thau, eds.), pp.~381--386, Boston, MA: Springer US, 1985.

\bibitem{hammandspaaij}
M.~S. Hamm and R.~Spaaij, {\em The Age of Lone Wolf Terrorism}.
\newblock Columbia University Press, 2017.

\bibitem{gtd}
``{G}lobal {T}errorism {D}atabase.'' \url{https://www.start.umd.edu/gtd/}.
\newblock Accessed: 2019-07-03.

\bibitem{motherjones}
``{M}other {J}ones.''
  \url{https://www.motherjones.com/politics/2012/12/mass-shootings-mother-jones-full-data/}.
\newblock Accessed: 2019-06-22.

\bibitem{silva2019fame}
``Fame-seeking mass shooters in america: Severity, characteristics, and media
  coverage,'' {\em Aggression and Violent Behavior}, vol.~48, pp.~24 -- 35,
  2019.

\bibitem{tierney2017spotting}
M.~Tierney, ``Spotting the lone actor: {C}ombating lone wolf terrorism through
  financial investigations,'' {\em Journal of Financial Crime}, vol.~24, no.~4,
  pp.~637--642, 2017.

\bibitem{eshgi2017mathematical}
S.~{Eshghi}, G.~{Williams}, G.~B. {Colombo}, L.~D. {Turner}, D.~G. {Rand},
  R.~M. {Whitaker}, and L.~{Tassiulas}, ``Mathematical models for social group
  behavior,'' in {\em 2017 IEEE SmartWorld, Ubiquitous Intelligence Computing,
  Advanced Trusted Computed, Scalable Computing Communications, Cloud Big Data
  Computing, Internet of People and Smart City Innovation
  (SmartWorld/SCALCOM/UIC/ATC/CBDCom/IOP/SCI)}, pp.~1--6, 2017.

\bibitem{ellis2016analysing}
C.~Ellis, R.~Pantucci, J.~D. R.~V. Zuijdewijn, E.~Bakker, M.~Smith, B.~Gomis,
  and S.~Palombi., ``Analysing the processes of lone-actor terrorism: Research
  findings,'' {\em Perspectives on Terrorism}, vol.~10, no.~2, pp.~33--41,
  2016.

\bibitem{capellan2015lone}
J.~A. Capellan, ``Lone wolf terrorist or deranged shooter? a study of
  ideological active shooter events in the {U}nited {S}tates, 1970–2014,''
  {\em Studies in Conflict \& Terrorism}, vol.~38, no.~6, pp.~395--413, 2015.

\bibitem{becker2014explaining}
M.~Becker, ``Explaining lone wolf target selection in the united states,'' {\em
  Studies in Conflict \& Terrorism}, vol.~37, no.~11, pp.~959--978, 2014.

\bibitem{roungas2018future}
B.~{Roungas}, A.~{Verbraeck}, and S.~{Meijer}, ``The future of contextual
  knowledge in gaming simulations: A research agenda,'' in {\em 2018 Winter
  Simulation Conference (WSC)}, pp.~2435--2446, Dec 2018.

\bibitem{agrawal1994fast}
R.~Agrawal and R.~Srikant, ``Fast algorithms for mining association rules in
  large databases,'' in {\em Proceedings of the 20th International Conference
  on Very Large Data Bases}, VLDB '94, (San Francisco, CA, USA), pp.~487--499,
  Morgan Kaufmann Publishers Inc., 1994.

\bibitem{nijkamp2008economic}
``Economic valuation of biodiversity: A comparative study,'' {\em Ecological
  Economics}, vol.~67, no.~2, pp.~217 -- 231, 2008.
\newblock Special Section: Biodiversity and Policy.

\bibitem{parack2012application}
S.~{Parack}, Z.~{Zahid}, and F.~{Merchant}, ``Application of data mining in
  educational databases for predicting academic trends and patterns,'' in {\em
  2012 IEEE International Conference on Technology Enhanced Education (ICTEE)},
  pp.~1--4, Jan 2012.

\bibitem{nazeri2001experiences}
Z.~Nazeri, E.~Bloedorn, and P.~Ostwald, ``Experiences in mining aviation safety
  data,'' in {\em SIGMOD '01}, 2001.

\bibitem{sarath2013association}
K.~Sarath and V.~Ravi, ``Association rule mining using binary particle swarm
  optimization,'' {\em Engineering Applications of Artificial Intelligence},
  vol.~26, no.~8, pp.~1832 -- 1840, 2013.

\bibitem{danping2011data}
Z.~Danping and D.~Jin, ``The data mining of the human resources data warehouse
  in university based on association rule,'' {\em Journal of Computers},
  vol.~6, no.~1, pp.~1832 -- 1840, 2011.

\bibitem{hamalainen2008efficient}
W.~{Hämäläinen} and M.~{Nykänen}, ``Efficient discovery of statistically
  significant association rules,'' in {\em 2008 Eighth IEEE International
  Conference on Data Mining}, pp.~203--212, 2008.

\bibitem{gras2006discovering}
R.~Gras and P.~Kuntz, ``Discovering r-rules with a directed hierarchy,'' {\em
  Soft Computing}, vol.~10, pp.~453--460, Mar 2006.

\bibitem{lewis2015human}
J.~W. Lewis, ``The human use of human beings: Suicide bombing, technological
  innovation, and the asymmetry of modern warfare,'' {\em Global Politics
  Review}, vol.~2, no.~2, pp.~9--27, 2015.

\bibitem{lansdekilde2019radicalization}
L.~Lindekilde, F.~O’Connor, and B.~Schuurman, ``Radicalization patterns and
  modes of attack planning and preparation among lone-actor terrorists: an
  exploratory analysis,'' {\em Behavioral Sciences of Terrorism and Political
  Aggression}, vol.~11, no.~2, pp.~113--133, 2019.

\bibitem{hubert1976general}
L.~J. Hubert and J.~R. Levin, ``A general statistical framework for assessing
  categorical clustering in free recall,'' {\em Psychological Bulletin},
  vol.~83, no.~6, pp.~1072--1080, 1976.

\end{thebibliography}
\bibliographystyle{ieeetr}

\newpage

\begin{appendices}
\section{Mathematical Interpretation of the \textit{A-priori} Algorithm}
\label{appendix:apriori}

Let $\mathcal{S}$ be a set of $d$ items, i.e. $\mathcal{S}={s_i: i=1,2,\dots,d}$, and $\mathcal{T}$ be a set of $n$ transactions in a database, i.e. $\mathcal{T}={t_i: i=1,2,\dots,n}$. An association rule is of the form $\mathcal{A} \rightarrow \mathcal{B}$ where $\mathcal{A}$ and $\mathcal{B}$ are mutually exclusive subsets of $\mathcal{S}$. The width ($w$) of $j^{th}$  transaction is defined as the number of items in the related transaction, that is, 
\begin{equation}\label{eq:width}
 w_j = \sum_{i=1}^d {\mathbbm{1}\{s_i \in T_j\}},
\end{equation}
\noindent with $\mathbbm{1}\{\cdot\}$ denoting the indicator function of the event defined in the parentheses, i.e.

\[
    \mathbbm{1}\{s_i \in T_j\}=
\begin{cases}
    1 & \text{if } s_i \in T_j\\
    0,              & \text{otherwise}.
\end{cases}
\]

 The calculation of {\it support\/} for an itemset $\mathcal{A}$ is defined as

\begin{equation}\label{eq:support}
\sigma_{\mathcal{A}} = \frac {\sum_{j=1}^{n}{{\prod_{s_i \in \mathcal{A}} {\mathbbm{1}\{s_i \in T_j\}}}} } {n}.
\end{equation}

\noindent Note that this is the fraction of transactions that demonstrate all the characteristics of itemset $\mathcal{A}$. If the frequency ratio of itemset $\mathcal{A}$ surpasses a given threshold, the itemset is considered to be frequent enough to be a candidate for an association generalization.

Association rules are constructed through mutually exclusive subsets of frequent itemsets. For a frequent itemset $\mathcal{A}$, an association rule is of the form $\mathcal{X} \rightarrow \mathcal{Y}$ where $\mathcal{X} \cup \mathcal{Y} = \mathcal{A}$ and $\mathcal{X} \cap \mathcal{Y} = \emptyset$, that is, the rules are constructed among all nonempty mutually exclusive subsets of $\mathcal{A}$. For the association rule of the form $\mathcal{X} \rightarrow \mathcal{Y}$, the {\it confidence\/} ($\gamma_{\mathcal{X}\mathcal{Y}}$) of a rule defines the frequency that, given a transaction contains the items in $\mathcal{Y}$, the transaction also contains the items in $\mathcal{X}$, and it is given as in Eq. \ref{eq:confidence}. A rule should hold at a certain confidence threshold. 
\begin{equation}\label{eq:confidence}
\gamma_{\mathcal{X}\mathcal{Y}} = \frac {\sum_{j=1}^n {\mathbbm{1}\{\mathcal{X} \in T_j\}} \cdot  {\mathbbm{1}\{\mathcal{Y} \in T_j\}} } {\sum_{j=1}^n  {\mathbbm{1}\{\mathcal{Y} \in T_j\}} }.
\end{equation}

The {\it lift\/} ($\lambda_{\mathcal{X}\mathcal{Y}}$), also called interestingness factor, is defined as a test measure that the posterior probability of the associations is higher than the prior probability (Eq. \ref{eq:lift}). A lift value higher than 1 indicates a strong dependence.
\begin{equation}\label{eq:lift}
\lambda_{\mathcal{X}\mathcal{Y}} = \frac {\sigma_{\mathcal{X}\mathcal{Y}} } {\sigma_{\mathcal{X}} \cdot \sigma_{\mathcal{Y}}} = \frac {\sum_{j=1}^n {\mathbbm{1}\{\mathcal{X} \in T_j\}} \cdot  {\mathbbm{1}\{\mathcal{Y} \in T_j\}} }  {\sum_{j=1}^n {\mathbbm{1}\{\mathcal{X} \in T_j\}} \cdot  \sum_{j=1}^n {{\mathbbm{1}\{\mathcal{Y} \in T_j\}} } }.
\end{equation}

\subsection{Statistical Significance of Association Rules}\label{appendix:signi}

For a rule $\mathcal{X} \rightarrow \mathcal{Y}$, this test uses normal approximation of binomial distribution. The support of $X$ ($\sigma_{\mathcal{X}} $) and the support of $\mathcal{Y}$ ($\sigma_{\mathcal{Y}} $) signify the probability of $\mathcal{X}$ and $\mathcal{Y}$ appearing in a random transaction, respectively. If $\mathcal{X}$ and $\mathcal{Y}$ are independent, $\mathcal{X} \cup \mathcal{Y}$ occurs in a transaction with probability $\sigma_{\mathcal{X}}\sigma_{\mathcal{Y}} $. The number of rows that contain $\mathcal{X}$  or $\mathcal{Y}$ is a binomial random variable. Then, given the transaction size of $n$, the number of transactions that contain $\mathcal{X} \cup \mathcal{Y}$ is  $n\sigma_{\mathcal{X}}\sigma_{\mathcal{Y}}$ where the variance is $n\sigma_{\mathcal{X}}\sigma_{\mathcal{Y}}(1-\sigma_{\mathcal{X}}\sigma_{\mathcal{Y}})$. With Chebyshev's inequality and the normal approximation to the binomial distribution, we test the hypothesis that $\mathcal{X}$ and $\mathcal{Y}$ are independent using the following formula:
\begin{equation}\label{eq:statistical}
P\Bigg(-K \leq \frac{\sum_{j=1}^n {\mathbbm{1}\{\mathcal{X} \in T_j\}} \cdot  {\mathbbm{1}\{\mathcal{Y} \in T_j\}} - n\sigma_{\mathcal{X}}\sigma_{\mathcal{Y}}}{\sqrt{n\sigma_{\mathcal{X}}\sigma_{\mathcal{Y}}(1-\sigma_{\mathcal{X}}\sigma_{\mathcal{Y}})}}  \leq K \Bigg) \geq 1-\frac{1}{K^2}.
\end{equation}

In this formula, $K \leq 2$, refers to the significance level of $p=0.05$. For testing multiple rules, we apply the Bonferroni adjustment. If m rules are tested, then $\sqrt(m)K$ must be used instead of $K$ for the same significance level.

\subsection{Association Rule Chains and R-Rules}\label{appendix:rrules1}

Let $R$ be an association chain of the form $R' \rightarrow R''$, where $R'$ is the set of left-hand sides and $R''$ is the set of right-hand sides of every rule involved in the chain. For example, the association chain $\mathcal{X} \rightarrow \mathcal{Y} \rightarrow \mathcal{Z}$ can be rewritten as $(\mathcal{X} \rightarrow \mathcal{Y}) \rightarrow (\mathcal{Y} \rightarrow \mathcal{Z})$ . Hence, for this rule $R'={\mathcal{X},\mathcal{Y}}$ and $R''={\mathcal{Y},\mathcal{Z}}$. Then, the cohesion of any association chain $R$ is defined as 
\begin{equation}\label{eq:cohesion1}
c(R) = \Bigg( c(R') \cdot c(R^{''}) \cdot  \prod_{a_i \in R', a_j \in R''} c(a_i,a_j) \Bigg),
\end{equation}
with 
\begin{equation}\label{eq:cohesion2}
c(R') = \prod_{i=1,...,r-1; j=i+1,...,r} c(a_i,a_j), a_i,a_j \in R',
\end{equation}
and
\begin{equation}\label{eq:cohesion3}
c(R'') = \prod_{i=1,...,r-1; j=i+1,...,r} c(a_i,a_j), a_i,a_j \in R''.
\end{equation}

Here, $r$ is the number of rules in the association, $a_i$ and $a_j$ are elements of sets $R, R'$ and $R''$, $c(R' )$ and $c(R'' )$ are the cohesion value of the rules in sets $R'$ and $R''$. The minimum coherence threshold is predetermined as 0.7, similar to the confidence value. As an additional robustness criterion for an association chain of the form $\mathcal{X} \rightarrow \mathcal{Y} \rightarrow \mathcal{Z}$, a minimum confidence threshold of 0.7 for $P(YZ|X)$ is also employed.

\end{appendices}

\end{document}